\documentclass{aa}
\usepackage{graphicx}

\begin{document}

\title{On the massive stellar population  of the Super Star Cluster \object{Westerlund~1}
\thanks{Based on observations collected at the European Southern
Observatory, La Silla, Chile (ESO 67.D-0211 \& 69.D-0039)}}
\author{J.~S.~Clark\inst{1}
\and I.~Negueruela\inst{2}
\and P.~A.~Crowther\inst{3}
\and S.~P.~Goodwin\inst{4}
}

\offprints{J. S. Clark, \email{jsc@star.ucl.ac.uk}}

\institute{ Department of Physics and Astronomy, University College London,
Gower Street, London, WC1E 6BT, England, UK
\and
Dpto. de F\'{\i}sica, Ingenier\'{\i}a de Sistemas y Teor\'{\i}a de
la Se\~{n}al, Universidad de Alicante, Apdo. 99, E03080 Alicante, Spain
\and
Department of Physics \& Astronomy, University of Sheffield, Sheffield, S3 7RH, England, UK
\and Department of Physics and Astronomy, University of Wales, Cardiff,
 CF24 3YB, Wales, UK.  }

\date{Received    / Accepted     }

\abstract{ We present new spectroscopic and photometric observations of the young Galactic open cluster
\object{Westerlund 1} (Wd~1) that reveal a unique population of massive evolved stars.  We
identify $\sim$200
cluster members  and present spectroscopic classifications for $\sim$25\% of
these. We find that all stars so classified are unambiguously post-Main Sequence objects, consistent with an
 apparent lack of an identifiable Main Sequence in our photometric data to V$\sim$20. We are able to
identify rich populations  of Wolf Rayet stars, OB supergiants and
short lived transitional objects. Of these, the latter group
consists of both hot (Luminous Blue Variable and  extreme B
supergiants) and cool (Yellow Hypergiant and Red Supergiant)
objects - we find that half the known Galactic population of  YHGs
resides within Wd1. We obtain a  mean  $V-$M$_{\rm V} \sim$25~mag
from the cluster Yellow Hypergiants, implying  a Main Sequence
turnoff at or below $M_{\rm V} =-5$ (O7 V or later). Based solely
on the masses inferred for the  53 {\em spectroscopically}
classified stars, we determine an absolute  minimum mass of $\sim
1.5 \times 10^3$M$_{\odot}$ for Wd~1. However, considering the
complete photometrically and spectroscopically selected cluster
population  and adopting a  Kroupa IMF we infer a likely mass for
Wd~1 of $\sim 10^5$M$_{\odot}$, noting that inevitable source
confusion and incompleteness are likely to render this an
underestimate. As such, Wd~1 is the most massive compact young
cluster yet identified in the Local Group, with a mass exceeding
that of Galactic Centre clusters such as the Arches and
Quintuplet. Indeed, the luminosity, inferred mass and compact
nature of Wd~1 are comparable with those of Super Star Clusters -
previously identified only in external galaxies - and is
consistent with expectations for a Globular Cluster progenitor.
 }

\titlerunning{The stellar content of Wd~1}

\maketitle

\section{Introduction}

Massive stars play an important role in the ecology of galaxies,
providing a major source of ionising UV radiation, mechanical
energy and chemical enrichment (Smith \cite{linda}). However,
serious gaps in our understanding of massive stars exist
throughout their lifecycle, in large part  due to  the rapid
evolution - and hence rarity - of such objects. For example,  the
lack of accurate observational constraints on the metallicity
dependant mass loss rates of such stars  - particularly for short
lived phases such as Luminous Blue Variables (LBVs) and Yellow
Hypergiants (YHGs) - restricts our ability to follow their
post-Main Sequence (MS)  evolution. In particular, currently we
cannot predict what path a  star of given initial mass will follow
as it evolves from the Main Sequence through the Wolf Rayet star
(WR) phase to supernova (e.g. the 'Conti' scenario; Conti
\cite{conti}, Maeder \& Conti \cite{mc}). Clearly, the most direct
way to address such problems is to identify and study massive
stars within clusters, since this enables   us to study a coeval
population at a  uniform metallicity, where the MS turnoff mass
and hence post-MS progenitor mass may be accurately determined
(e.g. Massey et al. \cite{massey01}).

Moreover, the study of clusters is important to the wider galactic evolution, since it
 is generally thought  that  massive stars form preferentially (perhaps
exclusively) in star clusters (e.g. de Grijs \cite{dG}). An
extreme example of this phenomenon is observed in
 starburst galaxies,  with star formation occurring
in   Super Star Clusters (SSCs), which are inferred to be
several orders of magnitude more massive than
galactic open clusters. Such objects  may  have very different basic properties from their smaller relatives,
 with  many authors  arguing  for different kinds of top-heavy
mass distributions (e.g. \object{M82-F}; Smith \& Gallagher \cite{sg}).
 The relevance of such a possibility is clear when one
considers that a large fraction of the stellar population of galaxies
may have formed in starburst episodes in the distant past. If such
episodes, in which very large numbers of stars form, are in any way
fundamentally different from the much more modest star formation episodes
that we observe in the present Milky Way, our modelling of
the history and evolution of galaxies could be profoundly biased.

Unfortunately, dense clusters of massive stars are rare
in  the galaxy.  The  Arches \& Quintuplet Galactic Centre clusters
 (with masses of $\sim$1$\times$10$^4$M$_{\odot}$;
Figer  \cite{figer04}) are the most massive known; however,
observations are hampered by their extreme reddening. Recently,
observations of the apparently unremarkable cluster Westerlund 1
(henceforth Wd~1) have demonstrated that for over 4 decades the
astronomical community has possibly overlooked the presence of a
SSC in our own galaxy.

Discovered   by  Westerlund (\cite{westerlund61}), photometric (Borgman et
al. \cite{borgman}, Lockwood  {\cite{lockwood} and Koornneef
\cite{koornneef}) and spectroscopic (Westerlund \cite{westerlund87}; West87)
surveys   suggested the presence of a population of highly luminous  early
and late type supergiants. However, despite these observations
 \object{Wd~1} languished in relative obscurity throughout the past decade,
 due in large part  to the significant reddening  (A$_v \sim$12.9~mag. 
Piatti et al.  \cite{piatti98})
which made high resolution  spectroscopic observations difficult.

Motivated by the unusual radio and mid-IR properties of two
cluster members (Clark et al. \cite{clark98}), we have undertaken
an extensive program of  spectroscopic and photometric
observations of \object{Wd~1}. The first results of this program -
the discovery of an unexpected large population of WRs   and a new
LBV - have already been reported in Clark \& Negueruela
(\cite{clark02}, \cite{clark04}). Here we report on the remainder
of the observations, specifically their use in constraining the
properties of the population of massive post MS objects within
\object{Wd~1}.

The paper is therefore structured as follows. In Sect. 2 we
briefly describe the observations and the data reduction
techniques employed.  Sections 3 \& 4 are dedicated to the
presentation and analysis of both spectroscopic and photometric
datasets. Finally, we present an analysis of the global properties
of the cluster in Sect. 5 and summarise the results in Sect. 6.

\section{Observations \& Data Reduction}

Low resolution spectroscopy of cluster members over the
red/near-IR spectral region ($\sim$6000 to 11000{\AA}) was taken on
2001 June 23rd to 25th from the ESO 1.52 m telescope at La Silla
Observatory, Chile. The telescope
 was equipped with  the Boller \& Chivens spectrograph, the Loral \#38
camera
and the \#1 (night 1) and \#13 (nights 2 and  3) gratings, giving
dispersions of $\sim 5$ \AA/pixel and  $\sim 7$ \AA/pixel - leading to
resolutions of $\approx 11$\AA\ and $\approx 16$\AA\ - respectively.

Intermediate-resolution spectroscopy of the brighter cluster
members was carried out on 2002 June 7th, using the ESO Multi-Mode
Instrument (EMMI) on the 3.6 m New Technology Telescope (NTT) at
La Silla. The Red Medium Resolution mode was used. The instrument
was equipped with gratings \#6 and \#7 in the red arm and the new
$2048\times4096$ MIT CCD mosaic (in the $2\times2$ binning mode).
Grating \#7 covered the $\lambda\lambda$6310--7835\AA\, region at
a dispersion of $\sim0.8$\AA/pixel, while grating \#6 covered the
$\lambda\lambda$8225--8900\AA\, range at $\sim$0.4{\AA}/pixel.

Data pre-reduction was carried out with MIDAS software, while
image processing and reduction were accomplished with the {\em
Starlink} packages {\sc ccdpack} (Draper \cite{draper}) and {\sc
figaro} (Shortridge \cite{shortridge}).

Finally  photometric observations of the field containing
\object{Wd 1} were made with the SUperb-Seeing Imager 2
(SuSI2) direct imaging camera on the NTT; the 2 mosaiced
2k$\times$4k CCDs providing a 5.5$\times$5.5 arcmin field of view.
Broadband UBVRI\footnote{The U band frames produced  no useable
photometry of cluster members due the to high reddening towards
Wd~1 (Sect. 4) and so are not discussed further.}  images were
obtained in service time on 2001 August 21, with integration times
for individual frames ranging from 1200~s (U band) to 2~s (R \& I
bands). Given the significant interstellar reddening anticipated
for \object{Wd 1} a selection
 of nearby red Landolt (\cite{landolt}) standards was observed. Debiasing and flat fielding (both dome and
twilight flat field frames were obtained) were accomplished with
the {\sc Starlink} package {\sc kappa}. Final photometry was
kindly extracted for us by Peter Stetson, using the {\sc daophot}
package in IRAF\footnote{IRAF is distributed by the National
Optical Astronomy Observatories, which are operated by the
Association of Universities for Research in Astronomy, Inc., under
cooperative agreement with the National Science Foundation.}
(Stetson \cite{stetson}).

\section{Spectroscopic classification.}

\begin{figure*}
\resizebox{\hsize}{!}{\includegraphics[angle=0]{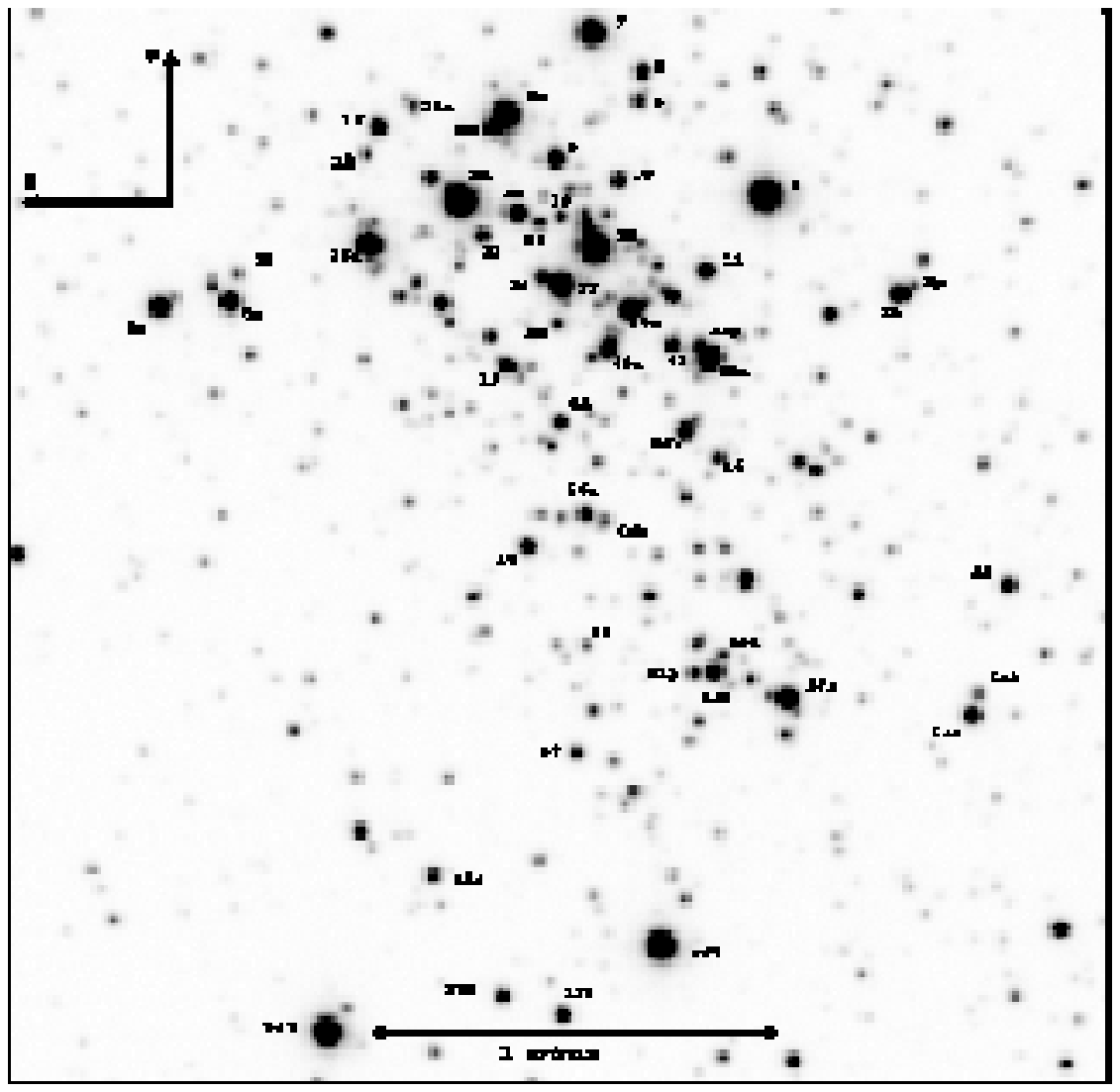}}
\label{Figure 1} \caption{Optical finding chart for
\object{Wd~1}. Several additional objects have also been indicated
where unclear on Fig. 1 of West87; \object{W31} has not been
labelled due to crowding, but is the blended object contiguous
with and to the north of, \object{W32}, while  W43a \& W42a are
the brighter, southernmost components of their respective blends.}
\end{figure*}

In total, 27 stars were observed at intermediate resolution
between 6500 and 7900{\AA} and 8200 and 8900{\AA}, and an
additional 26 stars  - including the 11 WRs identified in Clark \&
Negueruela (\cite{clark02}) - were observed at low resolution
between 6500 and 8900{\AA}. A finding chart for all
spectroscopically observed stars is presented in Fig. 1. Spectra
are presented in Figs. 2 (new WR candidates; Sect. 3.1), 3-7
(Super- and Hyper-giants; Sect. 3.2-3.5) and 8 (the unusual sgB[e]
star W9; Sect. 3.3.1).

The  6500 to 9000{\AA} window offers the prospect of spectral
classification of moderately reddened objects for which
traditional optical criteria are unavailable. Given the presence
of transitions from H\,{\sc i}, He\,{\sc i}  and numerous low -
e.g. Fe\,{\sc i}, Ca\,{\sc ii} and N\,{\sc i}  - and high - e.g.
 He\,{\sc ii}, C\,{\sc iii} and  N\,{\sc iv} - ionisation
species it would appear possible to construct an accurate
classification scheme. While authors have recognised the
importance of this window (e.g. Andrillat et al.
\cite{andrillat95}), comparatively few observations exist within
the literature, particularly  of high luminosity and/or early type
stars. We were therefore forced to adopt a twin approach for
spectral classification of our spectra - comparison to
observational data where possible (spectral type
mid-B$\rightarrow$G) supplemented by construction of a grid of
synthetic spectra where required (OB supergiants; Appendix A).

The results of the spectral classification are presented in Table 1, along with the original classification
by West87. Where possible we have adopted the nomenclature of West87; if an individual object
is resolved into two or more stars in our data we adopt the convention of retaining the designation of West87, with the
addition of the suffix {\em a} for the brightest component, {\em b} for the second brightest etc. We further present
the J2000 co-ordinates  and the BVRI broadband photometry for each object if available.

Note that while we have utilised a grid of synthetic spectra for the
classification of the OB star population of Wd~1 we {\em do not} claim
that these represent tailored analyses of individual stars. Given the
breadth and complexity of the spectra presented, quantitative analysis is
clearly beyond the scope of the current work, which aims to provide coarse
spectral classifications for the objects presented.

\subsection{The candidate Wolf Rayet Stars}

Clark \& Negueruela  (\cite{clark02}) reported  the serendipitous
discovery of 11 WR stars within  Wd1 in the low resolution 2001 ESO 1.52 m 
dataset, comprising  6 WN and 5 WC stars. Analysis of
the 2002 intermediate resolution NTT data reveals a further two
candidates within the cluster. These are W44 and W66 (=Candidates
L \&  M respectively, following the notation of Clark \&
Negueruela \cite{clark02}); 6500 to 7900{\AA} spectra for both
stars are  presented in Fig 2.

Turning first to W66, Fig. 2 reveals prominent blended emission in
He\,{\sc ii} 6560+C\,{\sc ii} 6578{\AA},  C\,{\sc ii}
6725-42+C\,{\sc iii} 6727-73{\AA}, C\,{\sc iii} 7037+He\,{\sc i}
7065+C\,{\sc iv} 7064{\AA} and C\,{\sc ii} 7236+C\,{\sc iii}
7210-12{\AA}. Difficulty in defining a continuum shortwards of
7000{\AA} prevents the accurate measurement of line strengths for
the first 3 blends, but we measure an
 EW$\sim -130 \pm$15{\AA} for the C\,{\sc ii} 7236+C\,{\sc iii} 7210-12{\AA} blend.
For the lowest excitation WC8 star (\object{WR 53}), the C\,{\sc iv} 
7065{\AA} and  C\,{\sc ii} 7236{\AA}
lines are of similar strength, while all  WC9 stars have C\,{\sc ii}
7236{\AA} $>>$  C\,{\sc iv} 7065{\AA}, indicating a classification of WC9 for W66.

W44 demonstrates emission in the  comparatively low ionisation lines of 
H$\alpha$
($\sim -30${\AA}), He\,{\sc i} 6678 ($\sim -8${\AA}) and 7065 ($\sim -14${\AA}), with no emission
from either higher (e.g. He\,{\sc ii}, N\,{\sc iii}-{\sc iv}  or C\,{\sc iii}-{\sc iv}) or lower
ionisation species (e.g.  Fe\,{\sc ii}). The spectrum in the
 8200 to 8900{\AA}  range appears  completely featureless 
(to a limit of  EW $\pm$ 0.5{\AA}) and hence is not
presented. Given the apparent magnitude of W44, combined with the strength of the emission lines,
 the lack of Paschen lines (emission and absorption) and low ionisation
species, we exclude a B emission line star classification - such as classical Be  
or sgB[e] star - and instead suggest
a WN identification.  The lack of  N\,{\sc iv} 7116{\AA}  precludes a classification earlier than WN9, with a
classification of WN9 rather than WN9ha more likely due to  the comparative strength of the
He\,{\sc i} lines. This is the first such star identified within Wd~1, although
 we note that the low resolution spectrum of W5 is possibly indicative of an
even later  WNL classification - this is detailed in Sect. 3.3.3.

\begin{figure}
\resizebox{\hsize}{!}{\includegraphics[angle=0]{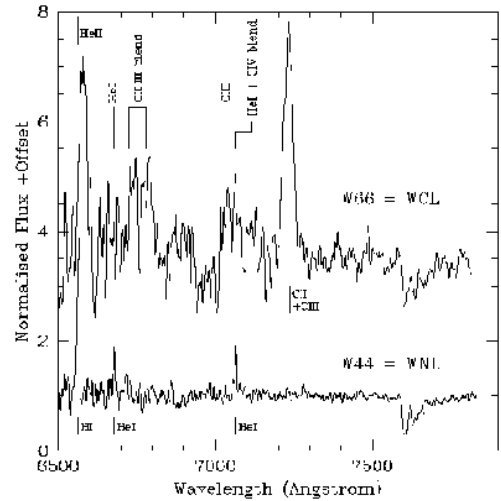}}
\label{Figure 1}
\caption{R band spectra of the 2 newly discovered WR stars in the cluster; \object{Candidate L} (=
\object{W44}),
a probable WN9 star and  \object{Candidate M} (=\object{W66}), a sixth WCL star (WC9).}
\end{figure}

\subsection{The late O \& early-B supergiants}

\begin{table*}
\begin{center}
\caption{Summary of the 53 cluster members with spectroscopic
classifications. Column 1 provides the notation of West87, while
Column 2 provides our additional notation if required. Columns 3
\& 4 provide the Equ J2000 co-ordinates, while broadband BVRI
photometric values are presented in Columns 5-8 where available.
Columns 9 \& 10 list spectral classifications from West87 and this
work, respectively. Classifications listed in italics were derived
from low resolution spectra. In general we find an encouraging
correspondence between the two works, although our classifications
appear to be systematically earlier for those stars classified as
late B-early G by West87. Importantly we confirm of the
classification  of West87's four YHG  candidates, while
identifying a further 2 candidates. Finally, we highlight one
important difference; we are unable to identify {\em any} Main
Sequence or giant stars, with {\em all} the OB stars in our study
being classified as Supergiants (W6, 10, 11, 28, 33, 70 \& 71
receiving revised luminosity classes). }
\begin{tabular}{cccccccccc}
\hline
W87  & ID    & RA (J2000) & Dec (J2000) & B & V & R & I &  Spectral &  Type \\ \cline{9-10}
     & (new) &            &             &   &   &   &   &  W87  & New \\
\hline
2   & 2a   &  16 46 59.71 & -45 50 51.1 & 20.4 & 16.69 & 14.23 & 11.73 & -    & O9.5 Ia - B0.5 Ia\\
4   &      &  16 47 01.42 & -45 50 37.1 & 18.7 & 14.47 & 11.80 & 9.15  & G0 Ia$^+$&  F2 Ia$^+$ \\
5   &      &  16 47 02.97 & -45 50 19.5 & 21.4 & 17.49 & 14.98 & 12.48 & -    & {\em WNL/early B Ia$^+$}   \\
6   &     &  16 47 03.04 & -45 50 23.6 & 22.2  & 18.41 & 15.80 & 13.16 & B1 V  &  {\em O9.5 Ia - B0.5 Ia}   \\
7   &      &  16 47 03.62 & -45 50 14.2 & 20.0 & 15.57 & 12.73 & 9.99  & A0 Ia & B5 Ia$^+$ \\
8   & 8a   &  16 47 04.79 & -45 50 24.9 & 19.9 & 15.50 & 12.64 & 9.89  & G0 Ia &  F5 Ia$^+$      \\
9   &      &  16 47 04.14 & -45 50 31.1 & 21.8 & 17.47 & 14.47 & 11.74 & B[e] & sgB[e] \\
10  &      &  16 47 03.32 & -45 50 34.7 &  -    &   -   &   -   &   -   & B0 III& {\em O9.5 Ia - B0.5 Ia}   \\
11  &      &  16 47 02.23 & -45 50 47.0 & 21.2 & 17.15 & 14.52 & 11.91 & B1 II & O9.5 Ia - B0.5 Ia   \\
12  & 12a  &  16 47 02.21 & -45 50 58.8 & 22.0  & 16.94 & 13.54 & 10.42 & A2 Ia$^+$&    A5 Ia$^+$ \\
13  &      &  16 47 06.45 & -45 50 26.0 & 21.1 & 17.19 & 14.63 & 12.06 & -    & OB binary/blend?   \\
14  & 14a  &  16 47 05.94 & -45 50 23.3 &  -    &   -   &   -   &   -   & -    &{\em OB+WN blend?}\\
15  &      &  16 47 06.63 & -45 50 29.7 & 22.8  & 18.96 & 16.38 & 13.75 & -    & {\em OB binary/blend?}   \\
16  & 16a  &  16 47 06.61 & -45 50 42.1 & 20.5 & 15.89 & 12.82 & 9.90  & A2 Ia$^+$& A2 Ia$^+$ \\
19  &      &  16 47 04.86 & -45 50 59.1 & 22.6  & 18.22 & 15.21 & 12.37 & -    & O9.5 Ia - B0.5 Ia\\
20  &      &  16 47 04.70 & -45 51 23.8 &    -  &   -   &   -   &  -    & M6 I  & $<${\em M6 I}\\
23  & 23a  &  16 47 02.57 & -45 51 08.7 & 22.1  & 17.85 & 14.91 & 12.07 & -    & O9.5 Ia - B0.5 Ia\\
24  &      &  16 47 02.15 & -45 51 12.4 & 23.0  & 18.71 & 15.96 & 13.24 & -    & OB binary/blend? \\
26  &      &  16 47 05.40 & -45 50 36.5 & 22.1  & 16.79 & 12.63 & 9.19  & M2 I  & $<${\em M6 I}\\
28  &      &  16 47 04.66 & -45 50 38.4 & 20.9 & 16.87 & 14.26 & 11.64 & B0 II & O9.5 Ia - B0.5 Ia \\
29  &      &  16 47 04.41 & -45 50 39.8 & 22.6  & 18.66 & 16.02 & 13.38 & -    & {\em O9.5 Ia - B0.5 Ia}   \\
30  &      &  16 47 04.11 & -45 50 39.0 & 22.4  & 18.45 & 15.80 & 13.20 & -    & OB binary/blend?\\
32  &      &  16 47 03.67 & -45 50 43.5 &   -   &   -   &   -   &   -   & G5 Ia &  F5 Ia$^+$   \\
33  &      &  16 47 04.12 & -45 50 48.3 & 20.0 & 15.61 & 12.78 & 10.04 & B8 I  & B5 Ia$^+$ \\
35  &      &  16 47 04.20 & -45 50 53.5 & 22.7 & 18.59 & 16.00 & 13.31 & -    & {\em O9.5 Ia - B0.5 Ia} \\
41  &      & 16 47 02.70  & -45 50 56.9 & 21.3  & 17.87 & 15.39 & 12.78 & -    & {\em OB binary/blend?}   \\
42  & 42a  &  16 47 03.25 & -45 50 52.1 &  -    &   -   &   -   &   -   & -    & {\em B5 Ia$^+$?}   \\
43  & 43a  & 16 47 03.54  & -45 50 57.3 & 22.8  & 18.05 & 15.22 & 12.26 & -    & O9.5 Ia - B0.5 Ia    \\
44  & L    & 16 47 04.20  & -45 51 06.9 & 22.6 & 18.86 & 15.61 & 12.52 & -    & WN9     \\
55  &      & 16 46 58.40  & -45 51 31.2 & 21.6 & 17.67 & 15.25 & 12.67 & -    & {\em O9.5 Ia - B0.5 Ia}   \\
56  &      & 16 46 58.93  & -45 51 48.8 & 21.7 & 17.46 & 14.81 & 12.15 & -    & {\em O9.5 Ia - B0.5 Ia}   \\
57  & 57a  & 16 47 01.35  & -45 51 45.6 & 20.7 & 16.54 & 13.83 & 11.13 & -    & B3 Ia       \\
60  &      & 16 47 04.13  & -45 51 52.1 & 22.8  & 18.50 & 15.96 & 13.28 & -    & O9.5 Ia - B0.5 Ia  \\
61  & 61a  & 16 47 02.29  & -45 51 41.6 & 21.2 & 17.16 & 14.62 & 12.01 & -    & O9.5 Ia - B0.5 Ia \\
    & 61b  & 16 47 02.56  & -45 51 41.6 & 22.7 & 18.59 & 16.00 & 13.31 & -    & OB binary/blend? \\
66  & M    & 16 47 03.96  & -45 51 37.5 &  -    & 19.79 & 16.85 & 13.96 & -    &  WC9       \\
70  &      & 16 47 09.36  & -45 50 49.6 & 21.2 & 16.88 & 14.10 & 11.29 & B8 Iab& B3 Ia \\
71  &      & 16 47 08.44  & -45 50 49.3 & 21.5 & 17.01 & 14.06 & 11.16 & B8 Iab& B3 Ia \\
72  & A    & 16 47 08.32  & -45 50 45.5 &  -    & 19.69 & 16.59 & 13.68 & -    & {\em WN4-5}  \\
237 &      &  16 47 03.09 & -45 52 18.8 & 22.8  & 17.49 & 13.00 & 9.40  &M6+ III& $<${\em M6 I}\\
238 &      &  16 47 04.41 & -45 52 27.6 & 21.4 & 17.47 & 14.98 & 12.45 & -    & {\em O9.5 Ia - B0.5 Ia}   \\
239 & F    &  16 47 05.21 & -45 52 25.0 & 21.7  & 17.86 & 15.39 & 12.90 &  -   & {\em WC9}  \\
241 & E    & 16 47 06.06  & -45 52 08.3 &  -    &   -   &   -   &   -   & -    & {\em WC9}  \\
243 &      & 16 47 07.55  & -45 52 28.5 &  -    &   -   &   -   &   -   & B2 Ia & LBV       \\
265 &      & 16 47 06.26  & -45 49 23.7 & 22.0  & 17.05 & 13.62 & 10.54 & G0 Ia$^+$&  F5 Ia$^+$   \\
 -  & B    & 16 47 05.35  & -45 51 05.0 &  -    & 20.99 & 17.50 & 14.37 & -    & {\em WN6-8}  \\
 -  & C    & 16 47 04.40  & -45 51 03.8 &  -    &   -   &   -   &   -   & -    & {\em WC8}  \\
 -  & D    & 16 47 06.24  & -45 51 26.5 &  -    &   -   &   -   &   -   & -    & {\em WN6-8}  \\
 -  & G    & 16 47 04.02  & -45 51 25.2 & 22.7  & 20.87 & 17.75 & 14.68 & -    & {\em WN6-8}  \\
 -  & H    & 16 47 03.91  & -45 51 19.9 & 23.0  & 18.55 & 15.46 & 12.46 & -    & {\em WC9}  \\
 -  & I    & 16 47 01.67  & -45 51 20.4 &  -    &   -   &   -   &   -   & -    & {\em WN6-8}  \\
 -  & J    & 16 47 00.89  & -45 51 20.9 &  -    &   -   &   -   &   -   & -    & {\em WN6-8}  \\
 -  & K    & 16 47 02.70  & -45 50 57.4 &  -    &   -   &   -   &   -   & -    & {\em WC8-9}  \\
\hline
\end{tabular}
\end{center}
\end{table*}

\begin{figure*}
\vspace*{10cm} \includegraphics{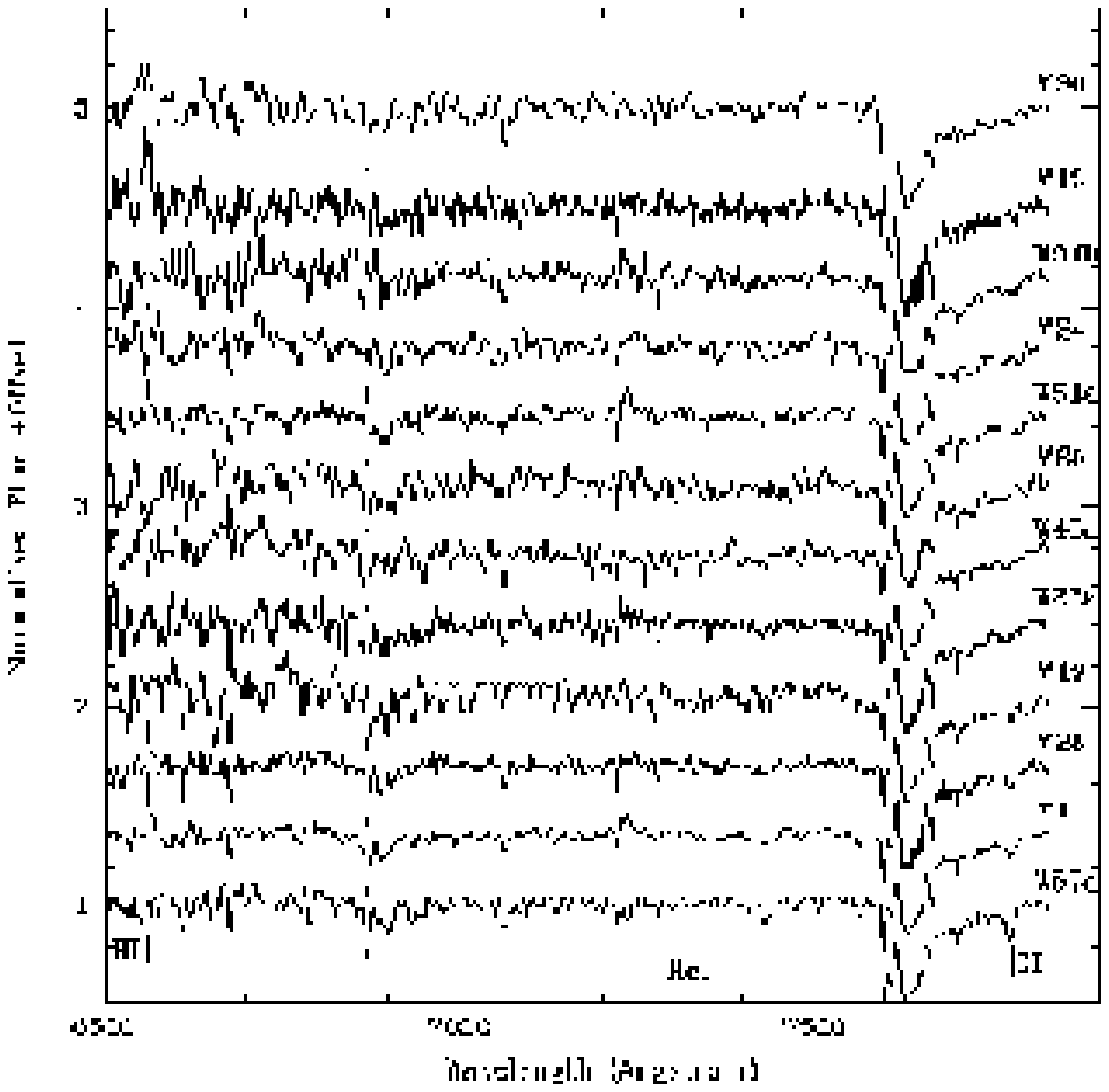} \includegraphics{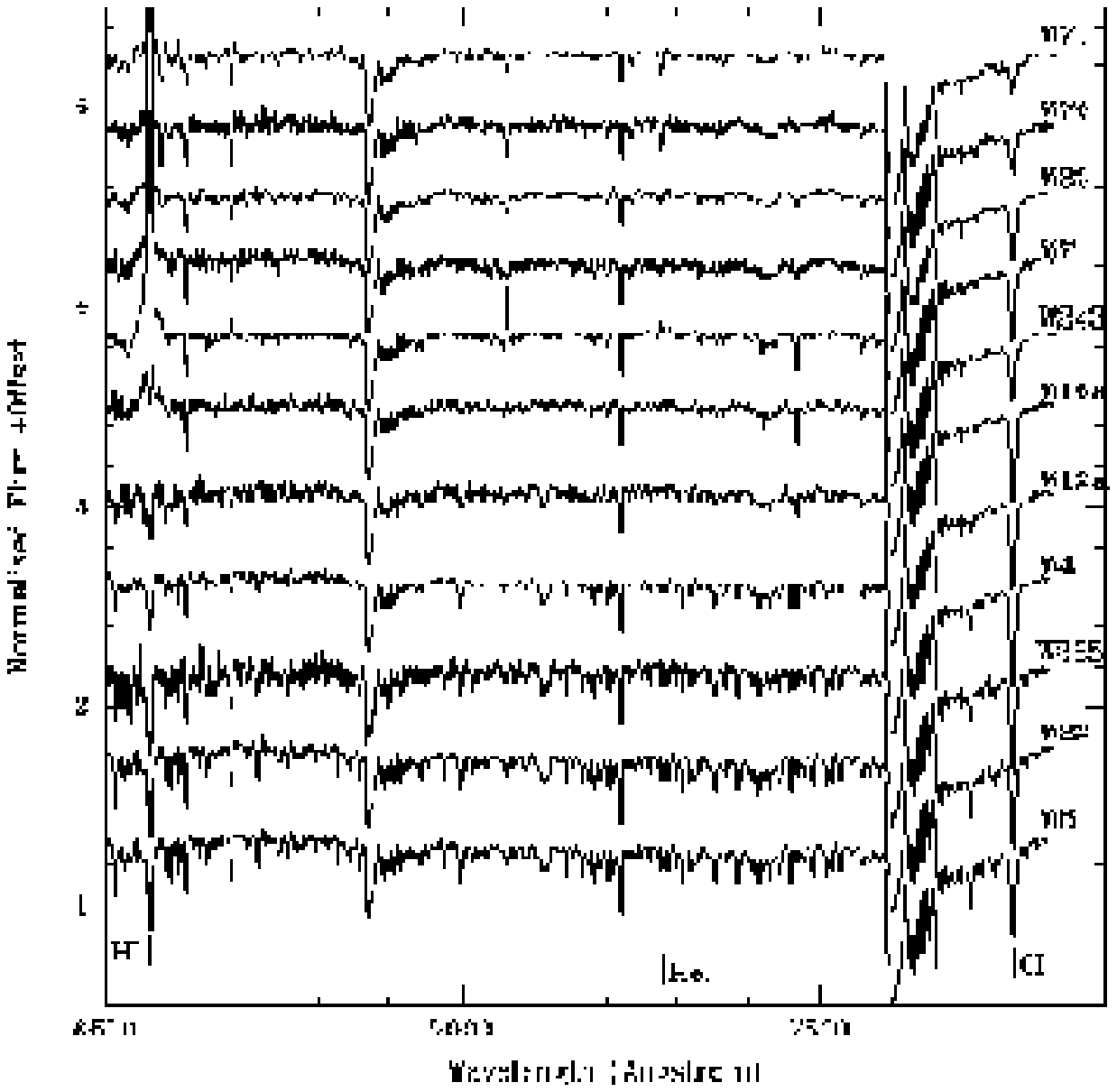}
\caption{Intermediate resolution R band (6400 to 7800 {\AA})
spectroscopy of O-F supergiant and hypergiant  members of Wd~1.
Prominent transitions  of H\,{\sc i}, He\,{\sc i} and O\,{\sc i}
are indicated. Left panel: O and mid B supergiants; we find no
evidence for He\,{\sc ii} absorption in the spectra. Right panel:
mid-B to
 F supergiants in \object{Wd~1}. Weak C\,{\sc ii} 6578/82~{\AA} absorption is visible
in the red wing of the H$\alpha$ line in the mid-B supergiants.
We tentatively identify the three prominent absorption features between
7420-7450~{\AA} in \object{W243} and \object{W16a} as N\,{\sc i} 7423.6,
7442.3 \& 7468.3~{\AA}. Weak Fe\,{\sc i} lines at 7112.2
and 6999.9~{\AA} also appear to be present in the spectrum of \object{W12a}
(A5Ia+) and later objects.
Possible identifications of the absorption features in the
mid-A and later spectral types at $\sim$7711~{\AA} and $\sim$7748~{\AA}
are  Fe\,{\sc ii} 7712 and Fe\,{\sc i}
 7748.3~{\AA} respectively.}
\end{figure*}

\begin{figure*}
\resizebox{\hsize}{!}{\includegraphics[angle=0]{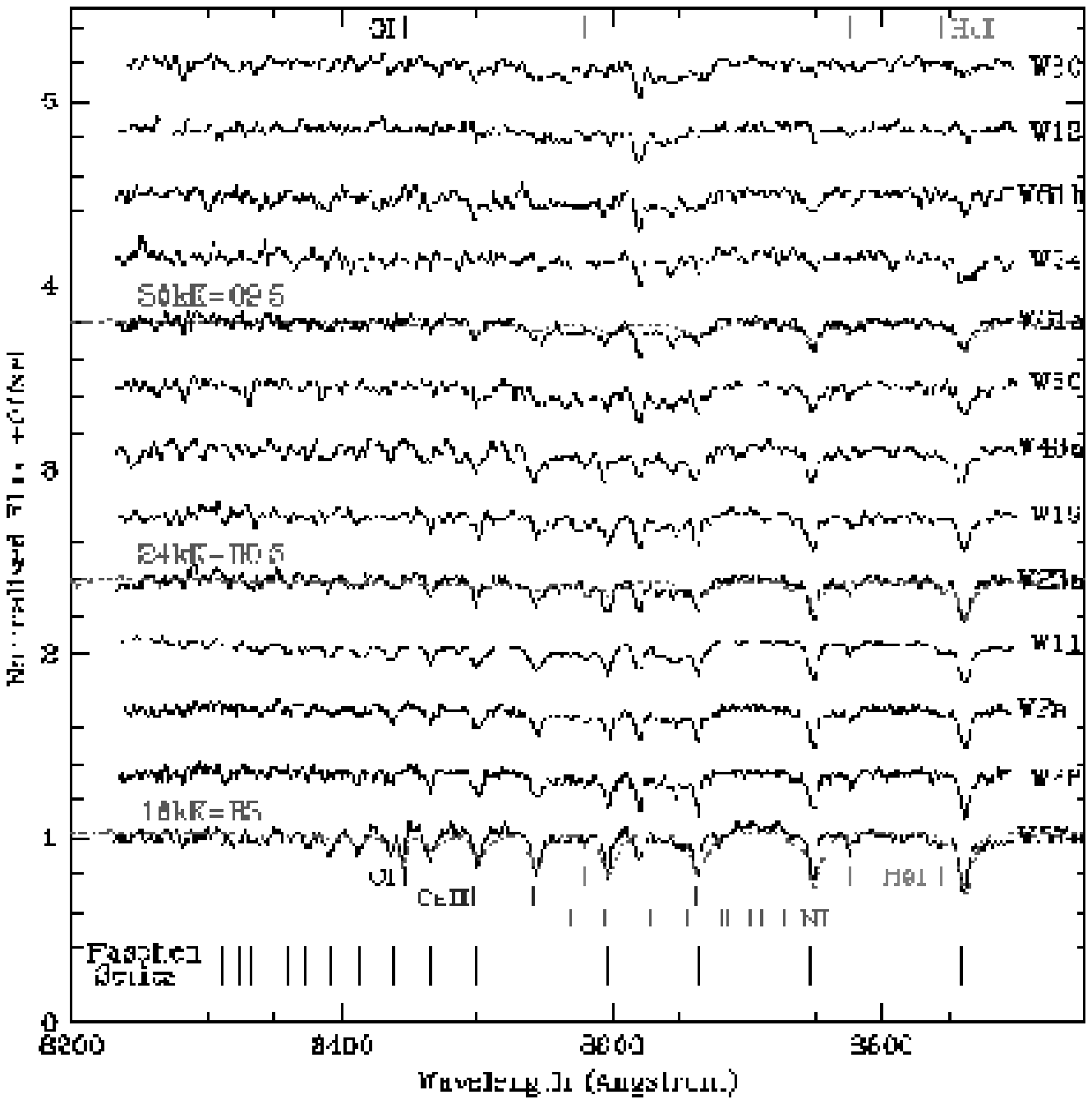}}
\label{Figure 1} \caption{Intermediate resolution I band (8200 to
8900{\AA}) spectroscopy (I): O and B supergiants in \object{Wd~1}.
We find the spectra of earliest stars to be almost featureless,
with the exception of the strongest Paschen series, while weak
He\,{\sc i} absorption appears later. In the  latest spectral
types ($\sim$B3) shown in this figure, weak absorption features of
O\,{\sc i} and N\,{\sc i} are also present. Note the feature at
$\sim$8620~{\AA} is a D.I.B. Representative synthetic spectra have
been overplotted on selected objects - note that these have {\em
not} been tailored to individual stars and hence we do not claim a
formal fit to these data; such a goal is left for a future paper.}
\end{figure*}

\begin{figure*}
\resizebox{\hsize}{!}{\includegraphics[angle=0]{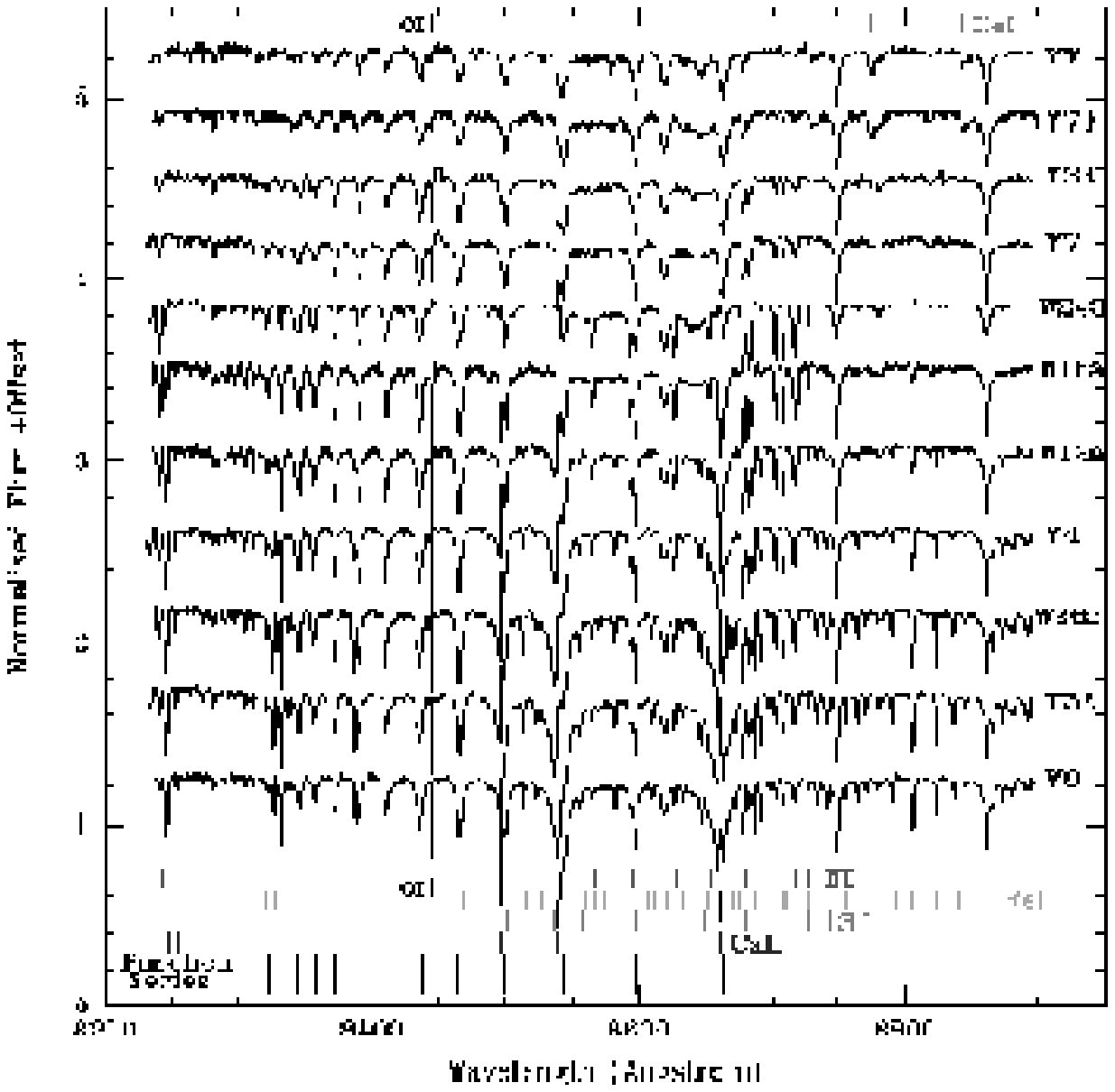}}
\label{Figure 1} \caption{Intermediate resolution I band (8200 to
8900 {\AA}) spectroscopy (II): mid-B to F supergiants in
\object{Wd~1}. All spectra are dominated by the H\,{\sc i} Paschen
series, while low  ionisation species such as Fe\,{\sc i},
Si\,{\sc i} and particularly N\,{\sc i} and Ca\,{\sc ii} come to
dominate the mid-A and later spectra. Despite their presence in
the later spectral types transitions of some low ionisation
species  such as e.g. Mg\,{\sc i} identifications are excluded for
reasons of clarity; we refer the reader to  Fig. 1 of Munari \&
Tomaselli (\cite{munari}) for more complete annotated spectra.
Note that we are unable to identify some of the (presumably low ionisation 
metallic) lines in the later spectral types, notably
the strong absorption feature at $\sim$8334{\AA} (blended with
Pa24). While not labelled, the feature at $\sim$8620~{\AA} is a
D.I.B.}
\end{figure*}

\begin{figure}
\resizebox{\hsize}{!}{\includegraphics[angle=0]{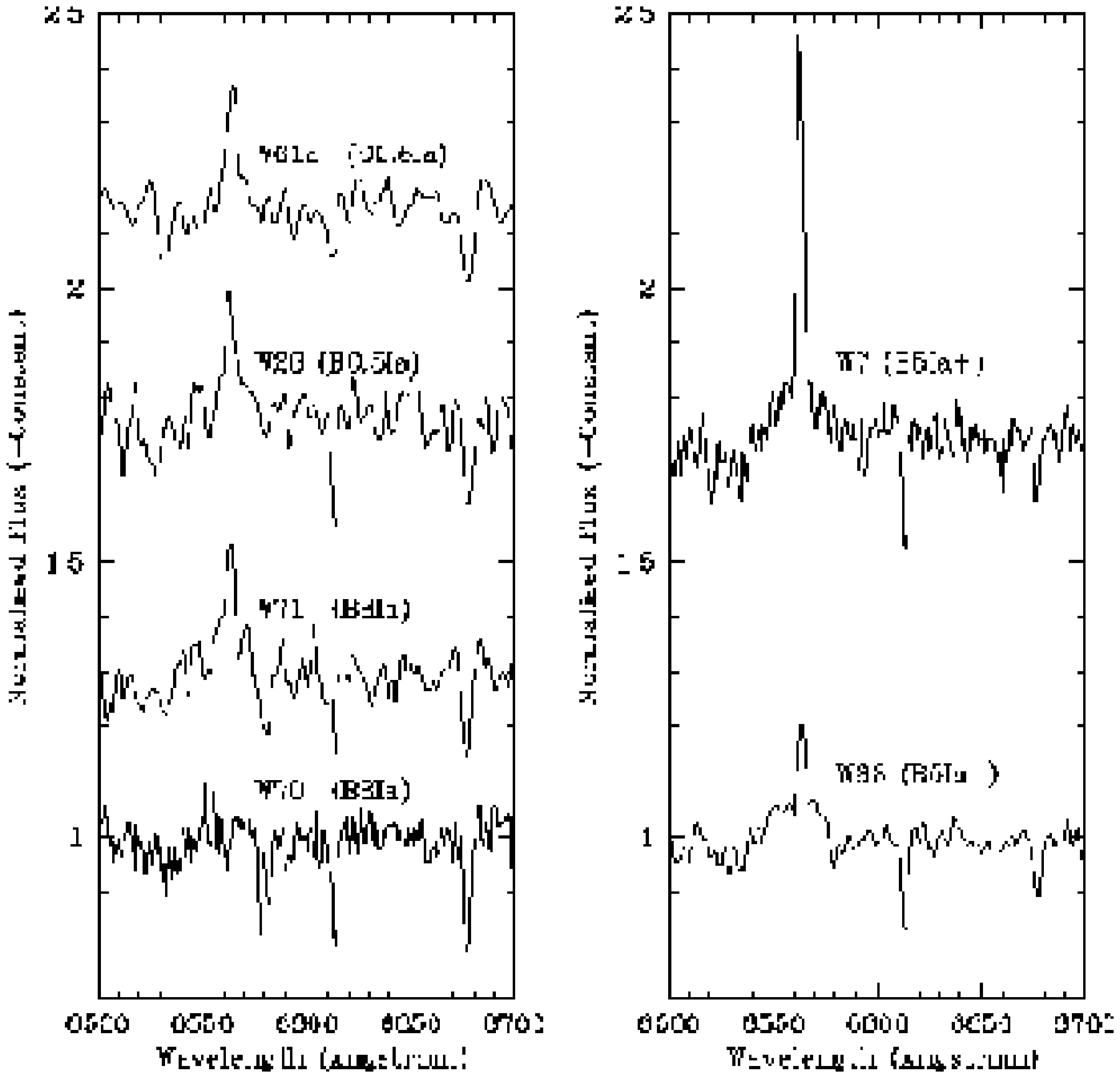}}
\label{Figure 1} \caption{Montage of the region of the spectrum
containing H$\alpha$ for a selection of the OB supergiants. The
left panel shows sample spectra for the O9.5 Ia to B3 Ia stars,
the presence of either infilling or H$\alpha$ emission indicating
a Ia luminosity classification. The right panel shows the
H$\alpha$ profiles for the two highly luminous mid-B hypergiants
W7 \& 33, note the narrow line superimposed on a broader plinth of
emission.}
\end{figure}

Intermediate resolution spectra of all 13 stars in Fig. 4 and W7,
33, 70 \& 71 in Fig. 5  fulfill the criteria for OB spectral types
defined in Appendix A and individual classifications are listed in
Table 1. The FWHM of the Paschen lines clearly satisfy the
criterion of Caron et al. (\cite{caron}) for supergiants, forming
a continuous sequence with the later A-F stars in Fig. 5, which we
demonstrate in Sect. 3.4 to also be super- or hyper-giants.

We find W57a, 70 \& 71 to have the latest spectral type - B3 Ia -
of the continuous sequence of OB supergiants present in the
sample. This classification  may be assigned on the basis of the
presence and strength of weak N\,{\sc i} and O\,{\sc i} 8446{\AA}
absorption. While W7 \& 33 apparently are of later spectral type,
their spectra demonstrate several unusual properties that suggest
they have more in common with the transitional objects such as the
LBV W243 and the YHGs; hence we defer discussion until Sect. 3.3.

The  remaining  spectra  display solely H\,{\sc i} and He\,{\sc i}
photospheric absorption lines, implying subtypes of  B1.5 Ia or
earlier (T$_{eff} \geq 20$kK). In addition, we find no signature
of the He\,{\sc ii} 8238{\AA}  line in any object, suggesting
 subtypes no earlier than O9.5 Ia. Further
subclassification is hampered by the poor S/N of most spectra, which precludes the use of the Pa16+C\,{\sc
iii}/Pa15 line ratio.

Consequently, we may only attempt additional classification based
on the strengths of the Paschen lines. Measuring the strengths of
the Paschen lines in the  observed (and synthetic) spectra
relative to  W70 \& 71 (B3 Ia model) suggests an approximate range
of spectral types for W61a, 60, 43a, 19, 23a, 11, 2a \& 28
(earliest $\rightarrow$ latest) of O9.5 Ia $\rightarrow$ B0.5 Ia
(30kK$\rightarrow$24kK). While the systematic weakening in line
strength  - resulting in the absence of higher Paschen series
lines - of stars such as W61a \& 60 compared to  W2a \& 28 is
clearly indicative of graduations in spectral type within this
group, we refrain from attempting any further subclassification
based on current data.

Finally, while the spectra of W30, 13, 61b \& 24 are almost completely devoid of photospheric features and hence
suggest spectral types  of O9Ia or earlier, the absence of He\,{\sc ii} 8238{\AA} apparently precludes
such a possibility. Instead we consider it likely that possible binarity or blending of 2 nearby objects leads
to the featureless spectra and hence refrain from a classification other than generic early OB supergiant for
these 4 objects.

Although the S/N for many spectra is rather poor, we find no evidence for a photospheric H$\alpha$ profile as
predicted by our synthetic spectra in the temperature range 10-34kK.
A montage of the region containing the H$\alpha$ line for those spectra with the largest  S/N is presented in
 Fig. 6; we find evidence for either infilling (W19, 23a, 24, 43a, 57a 60,
61b \& 70) or emission (W11, 13, 28, 30, 61a \& 71) in {\em all} spectra. Preempting  the results of
Sect. 4, in the absence of an accurate photometric determination of bolometric luminosity, we may use the presence of H$\alpha$
emission to infer a large luminosity and initial mass for the OB supergiants. Specifically, McErlean et al.
(\cite{mcerlean}) demonstrate that of their sample all the most luminous  (M$\sim$30-40M$_{\odot}$) B supergiants
demonstrate  H$\alpha$ emission.

Fig. 7 presents  a montage of the low resolution spectra of OB star candidates; where intermediate
resolution
spectra exist, these have  been overplotted. Given both the lower resolution
and S/N of the spectra, detailed classification  is difficult. Hence we simply choose to divide the spectra
into three morphological groups based on the presence of various photospheric features. Group 1 contains those stars
without an obvious Paschen series and consists of  W13, 14a, 15 \& 44. Group 2 contains those stars with the Paschen
series in absorption and consists of W2a, 5, 6, 10, 11, 23a, 24, 29, 43a, 55, 56 \& 238. Finally, group 3 consists of
those stars with both Paschen and the O\,{\sc i} 7774{\AA} blend in absorption, and consists of W7, 33, 41, 42, 70 \&
71.

Of the group 1 objects W44 and W13 have already been discussed. With an essentially featureless spectrum we suspect
that W15 is likely a binary or blend of one or more OB SGs. The strong N\,{\sc iv} emission seen in W14a is
unexpected since this would imply an early O supergiant  classification, at odds with the expected age of the cluster
(Sect. 5). However, we note that W14a appears to be  composed of 2 or more stars; hence an alternative
explanation
would be a blend containing at least one hitherto undetected WNE star. Clearly, future observations are required to
resolve this apparent inconsistency.

With a lack of O\,{\sc i} 7774{\AA} absorption, the group 2
objects appear to have earlier spectral types than $\sim$B3 Ia.
 Indeed all those objects for which intermediate resolution
observations are available are of spectral type O9.5 Ia to B0.5 Ia
- hence we assume a late O/early B supergiant classification for
the remaining objects, with the sole exception of W5, which,
together with W7 and 33, we discuss   in Sect. 3.3.3.

Of the group 3 spectra W41 \& 42 have only been observed at low resolution. W41 apparently shows weak O\,{\sc
i} absorption, although it is  coupled with an unexpectedly  weak Paschen series. H$\alpha$ is clearly seen in
absorption; the {\em only} such object in our dataset. Given the rather unusual nature of the spectrum we tentatively
suggest the spectrum  may be a blend of two or more stars. By contrast W42a has pronounced
 H$\alpha$ emission and appears  similar to the mid B supergiants W7 \& 33 (Sect. 3.3.3) and is discussed therein.

Finally, comparison of the two epochs of observations of W23a and 43a are suggestive of
variability, with H$\alpha$ emission conspicuous in 2001, but apparently  absent in 2002.

\begin{figure*}
\vspace*{10cm}
\includegraphics{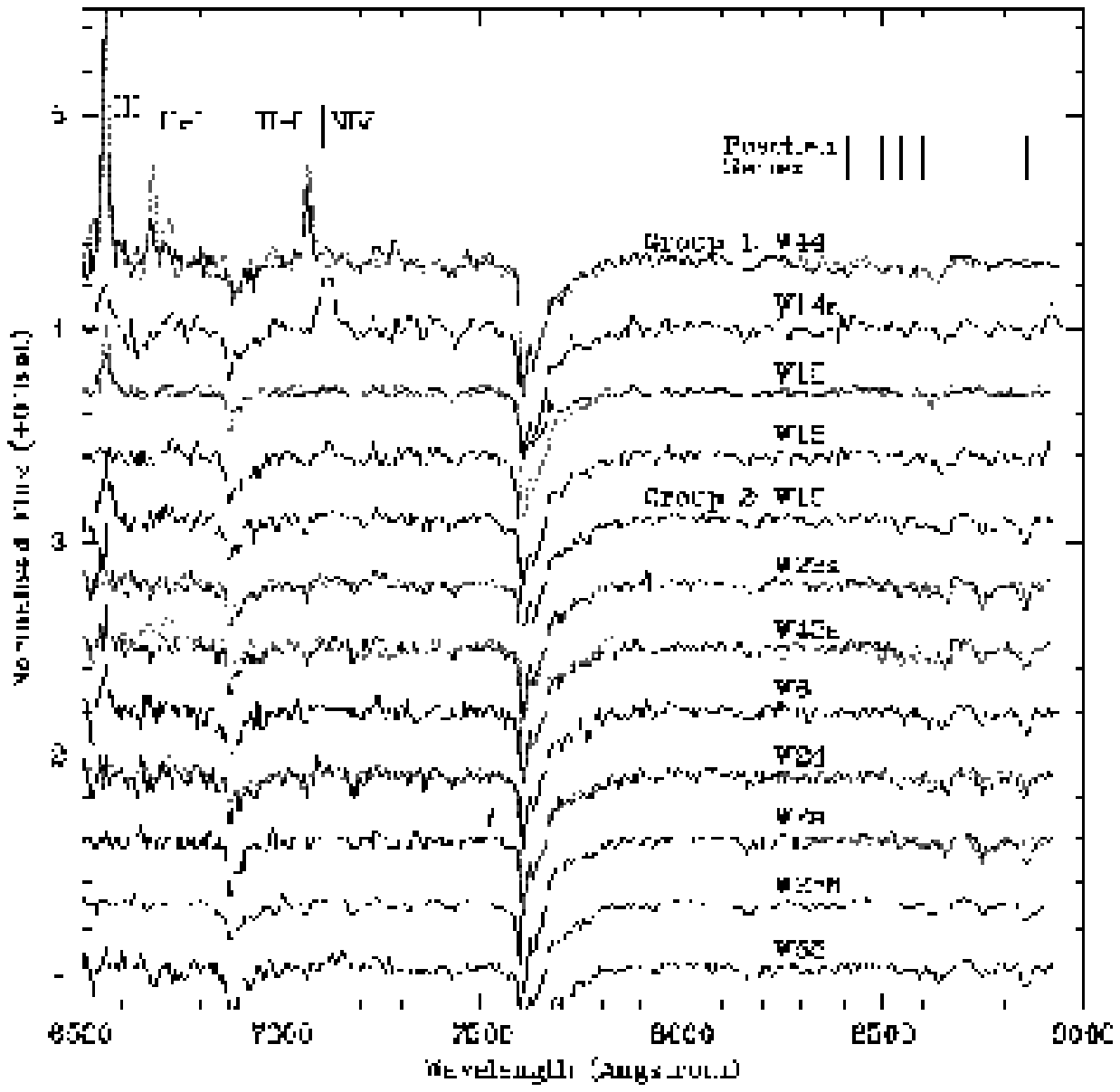}
\includegraphics{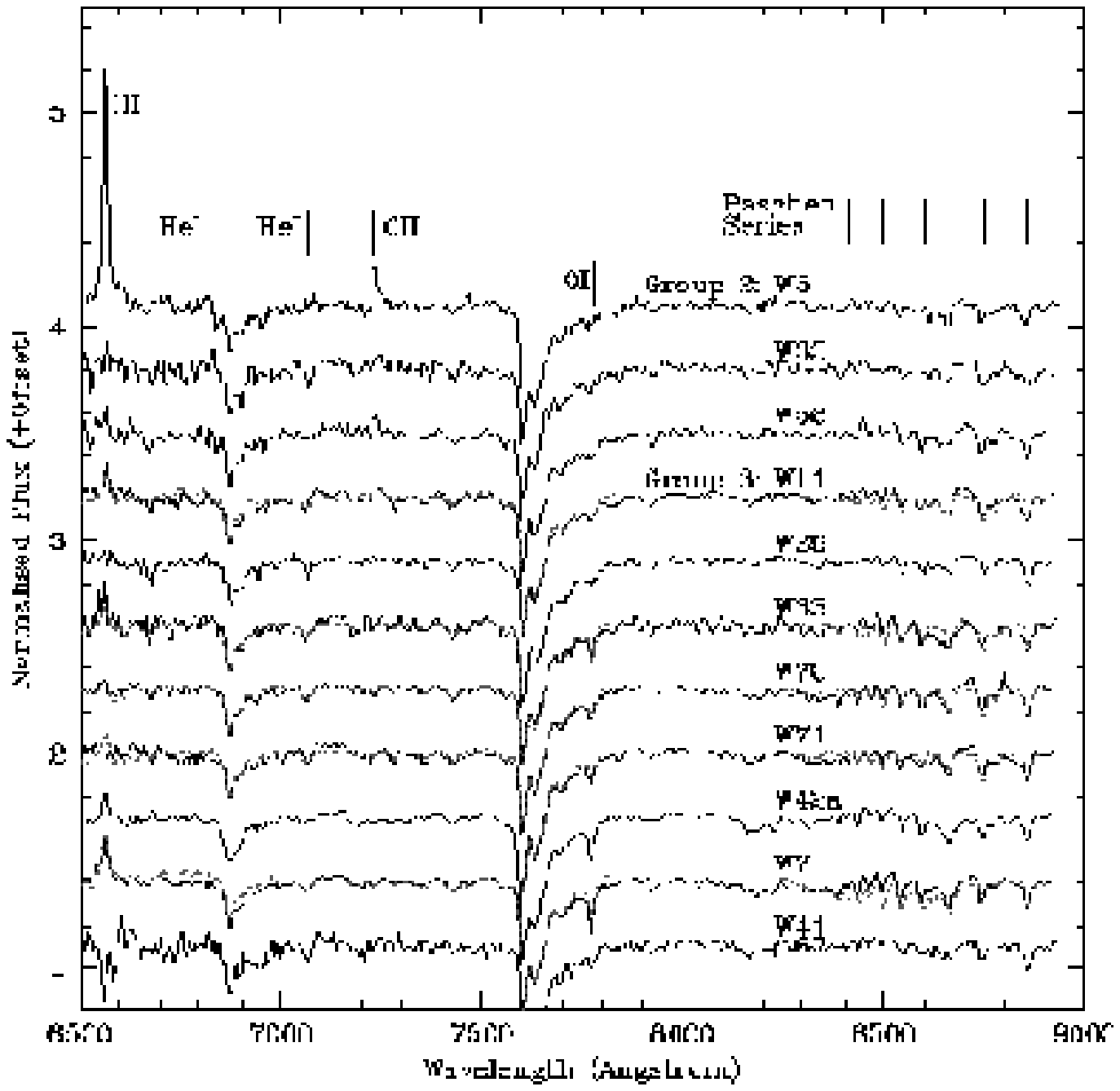}
\caption{Low resolution spectra of cluster OB supergiants (solid black lines). Where available high resolution
spectra (from Figs. 4-6) are smoothed and overplotted (dotted red lines) on the corresponding low
resolution spectra for comparison. Due to the low resolution an accurate determination of spectral type is impossible;
 hence we simply divide the spectra into 3 morphological classes based on the presence or otherwise of the Paschen
lines and the O\,{\sc i} 7774{\AA} feature; see Sect. 3.2 for
further details.}
\end{figure*}

\subsection{Anomalous B supergiants}

Amongst the observations of the OB supergiants within the cluster we find a significant number of
objects with spectra demonstrating unusual properties. In this subsection we detail these
individually.

\subsubsection{The sgB[e] star W9}

The spectrum of W9 is presented in Fig. 8 and is dominated by
H\,{\sc i}, He\,{\sc i} and low ionisation metallic emission
(Fe\,{\sc ii}, Ca\,{\sc i}, N\,{\sc i} and O\,{\sc i}), with
narrow, single peaked emission lines
(FWHM(H$\alpha$)=125kms$^{-1}$) and is notable for the complete
lack of {\em any} photospheric features and the strength of
certain emission features (e.g.
$W_{\lambda}$(H$\alpha$)=-520$\pm$30{\AA},$W_{\lambda}$(OI)=-48$\pm$2{\AA}).
To the best of our knowledge such an equivalent width for
H$\alpha$ is completely unprecedented for an early type emission
line star,  with the exception of the extreme systems
\object{SS433} (X-ray binary; e.g. Falomo et al. \cite{falomo})
and \object{$\eta$ Car} (LBV; e.g. Davidson et al.
\cite{davidson}). Radio observations (Clark et al. \cite{clark98},
Dougherty et al. \cite{dougherty}) reveal a strong, compact radio
source to be associated with W9, while mid IR observations
likewise show strong emission (F$_{10{\mu}m}$=50Jy), although the
emission mechanism (thermal dust or f-f/b-f) is still uncertain.
Finally, an ISO-SWS spectrum reveals the presence of [O\,{\sc iv}]
emission, implying an exciting source with a temperature in excess
of $\sim$80kK must be present (Clark et al. \cite{clark98}).

While a quantitative analysis of the combined dataset is beyond
the scope of this work, we note that the observational properties
of W9 are unique within Wd~1, and apparently amongst the wider
galactic population of early OB stars. While it fulfills the
classification  criteria for supergiant B[e] stars of Lamers et
al. (\cite{lamers}) we suspect that the emission line spectrum
arises at least partially  in the compact circumstellar envelope
rather than in a  stellar wind (Clark et al. \cite{clark98}). As
such it could closely resemble the unusual object \object{NaSt 1},
which Crowther \& Smith (\cite{crowther99}) propose to consist of
a compact ejection nebula surrounding  - and obscuring - an
exciting WNE object. Such an hypothesis would naturally explain
the [O\,{\sc iv}] emission present in W9, although one might
question the compact nature - and hence implied youth - of the
radio nebula when contrasted to other LBV nebulae, given the
requirement for the central star to evolve through both the LBV
and WNL phases.

Clearly, while  such an objection could be overcome by proposing a compact WR+LBV binary system, we also suggest an
alternative possibility, namely that W9 is the result of a recent merger event. Given the severely crowded nature of
the central regions of the cluster (Sect. 5.4) stellar collisions and interactions appear likely.
A recent ($\sim$10$^3$yr) interaction in  which the outer H mantle of one or both objects were lifted off to form the compact
radio nebula would reveal the hot interior layers of the star(s) so affected. This would then naturally
explain the high  excitation emission -  until the remnant relaxed to a new equilibrium structure on a thermal
timescale   ($\sim$10$^4$~yrs); a  scenario proposed by  Clark et al. (\cite{clark00}) to explain the stellar
and nebular morphology of the candidate  LBV \object{G25.5+0.2}.

\begin{figure*}
\vspace*{10cm}
\includegraphics{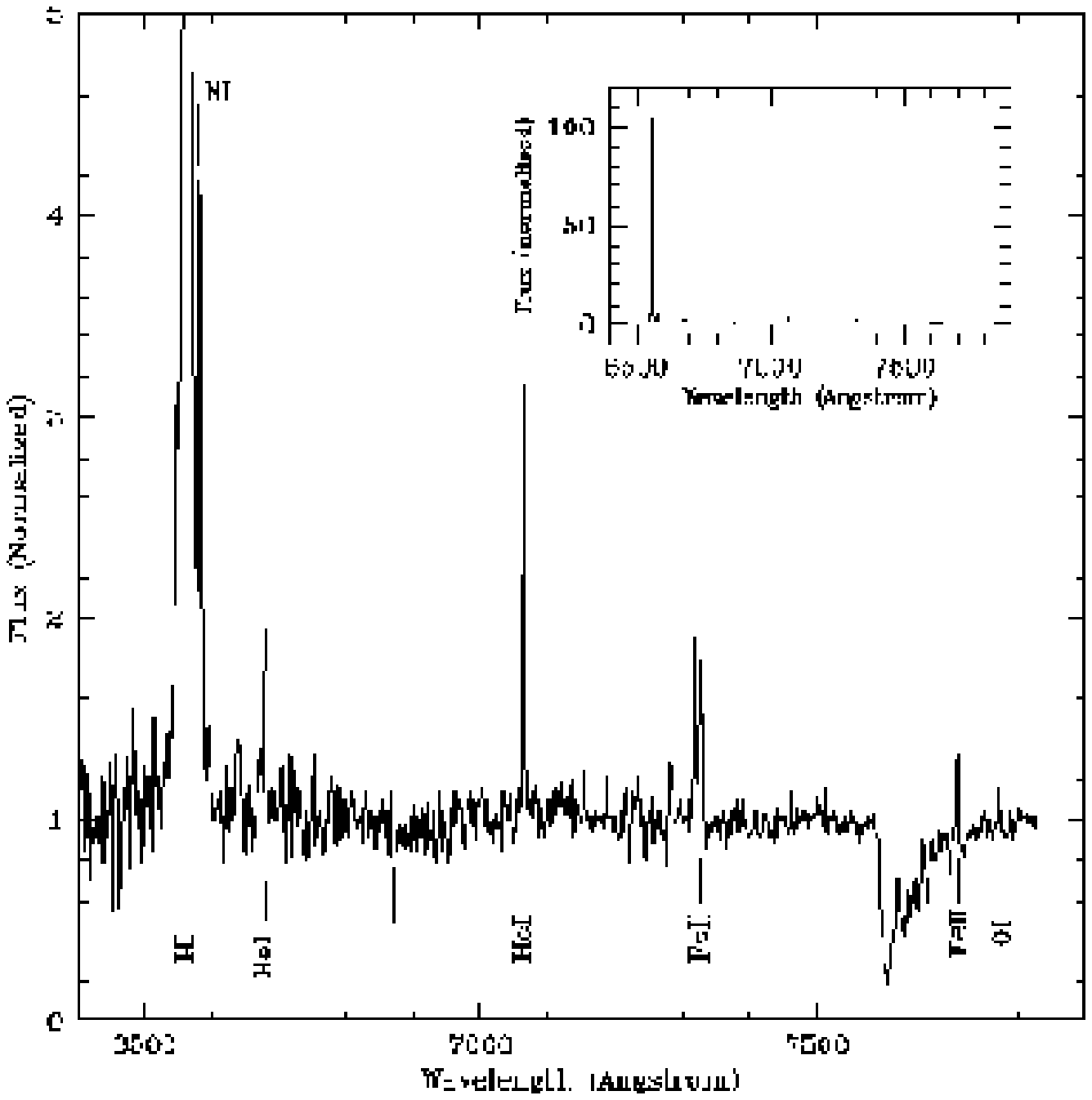}
\includegraphics{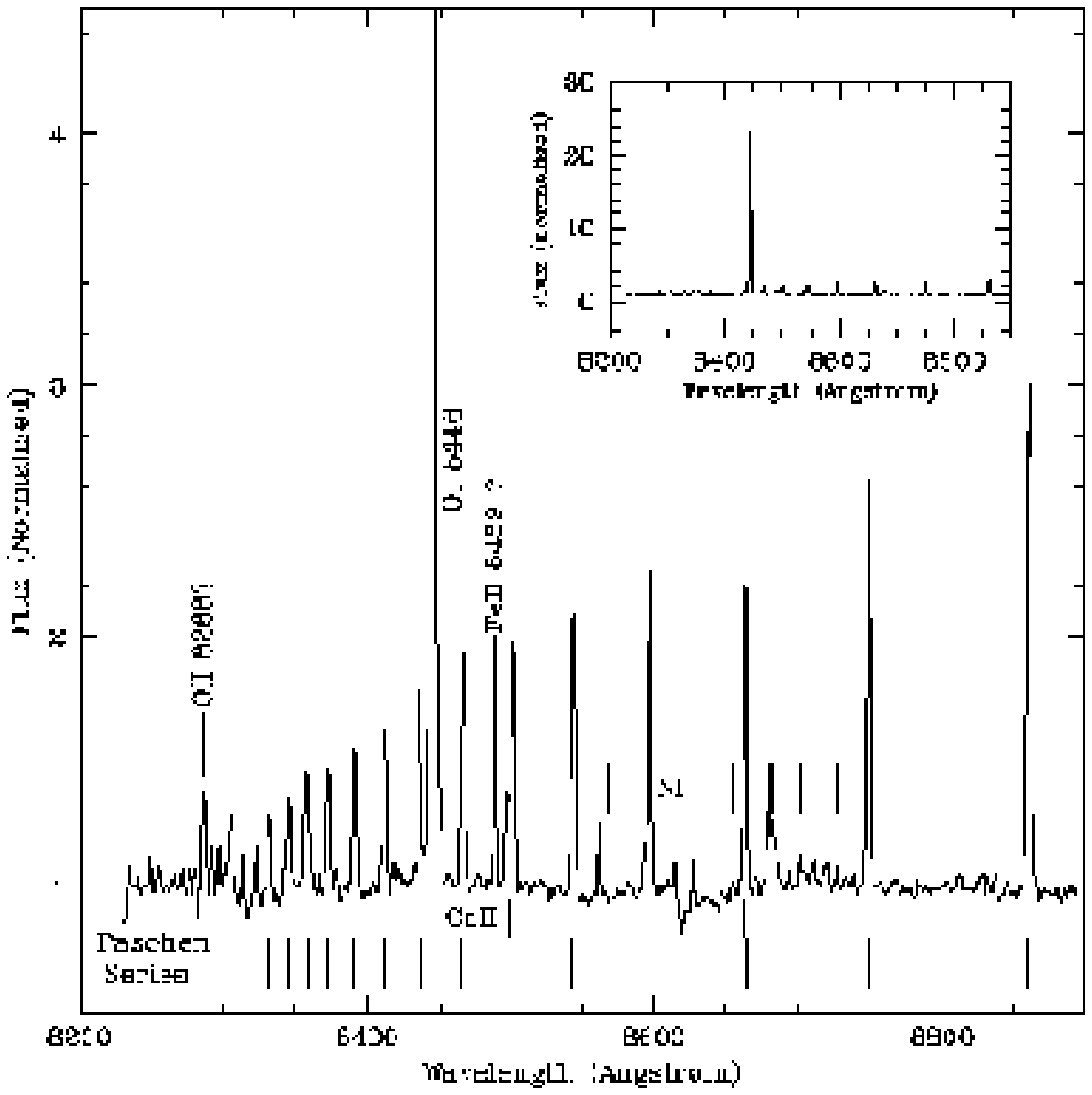}
\caption{R and I band spectra of  the sgB[e] star \object{W9}.
Note that we are unable to identify the  narrow emission features at
7134, 7233 \& 7280~{\AA}.}
\end{figure*}

\subsubsection{The Luminous Blue Variable W243}

Clark \& Negueruela (\cite{clark04}) present multi-epoch data that clearly demonstrate that W243
- initially classified as an early B supergiant by West87 - has evolved in
the intervening decades to
demonstrate a much cooler, A supergiant spectrum, albeit with anomalous emission features. The
overall spectrum closely resembles that of the luminous A supergiant \object{IRC +10 420} (e.g.
Oudmaijer \cite{oudmaijer}), with the notable exception of emission in the He\,{\sc i} 6678
and 7065{\AA} lines. Clark \& Negueruela (\cite{clark04}) advanced the suggestion that W243 was an
LBV, with the unusual composite spectrum resulting from  the formation of a pseudo-photosphere in a
dense stellar wind.

Subsequently, de Koter (2004, priv. comm.) suggested that as presented such an hypothesis
would predict significantly stronger H$\alpha$
emission (EW$\sim -100${\AA}) than observed.  However, the presence of He\,{\sc i} emission and
strong electron scattering wings (extending to $\pm$10$^3$kms$^{-1}$) in the H$\alpha$ emission
profile  apparently excludes a simple alternative explanation that W243 now has the physical
properties - and in particular a rather low mass loss rate - of a {\em bona fide} A supergiant.

Consequently, we suspect that a hybrid model may provide the correct explanation - {\em viz.} a
cooling of the underlying star to a late B spectral type - still sufficient to produce He\,{\sc i}
emission - coupled with an increase in mass loss rate to simulate a later spectral type.

\subsubsection{Extreme B supergiants W5, 7, 33 \& 42a}

Based on the  strength of the O\,{\sc i} 8446{\AA} and N\,{\sc i} features present in the intermediate resolution
spectra (Fig. 5), W7 \& 33 appear to be B5 supergiants. However, closer examination
shows a weak P Cygni  profile in the O\,{\sc i} 8446{\AA} line of both stars -
 to the best of our knowledge  absent from  any other mid-B supergiant, with the notable exception of the
2002 spectrum of the LBV W243.

Examination of the H$\alpha$ profiles also shows that they differ
from the weak P Cygni profiles expected for B supergiants, being
composed of a single narrow peak superimposed on a broad emission
base (Fig. 6). In this respect  they again resemble the LBV W243
(Fig. 3), while the broad emission component is observed for the
cluster member  YHG W16a (Sect. 3.4). Moreover, similar composite
Balmer line profiles are also obtained for the `B-supergiant LBVs'
studied by Walborn \& Fitzpatrick (\cite{walborn}). Consequently,
we  propose that both objects are to be found in  a shortlived,
transitional evolutionary state more closely related to the LBV
and YHG population of Wd~1 than the late O/early  B supergiant
population. Indeed, both stars  are a magnitude more luminous than
any other OB supergiant within the cluster,  with an observed V
band magnitude directly comparable to the YHG population.

Comparable objects might therefore be the highly luminous B
1.5a$^+$ hypergiants \object{HD~80077} (Carpay et al.
\cite{carpay}) and \object{HD~152236} (Van Genderen et al.
\cite{vG}), which are also characterised by strong H$\alpha$
emission due to mass loss rates  an $\sim$order of magnitude
greater than  Ia and Ib supergiants of similar spectral type (e.g.
Leitherer et al. \cite{leitherer}).

Of the stars for which only low resolution spectra exist, we find that W42a has a comparable morphology to both
W7 \&
33, suggesting a similar classification. W5 demonstrates both dramatic H$\alpha$  and also
C\,{\sc ii} $\sim$7236{\AA} emission, previously only identified in the WCL population. However, the lack of
C\,{\sc iii} 9705{\AA} emision (not shown in Fig. 7) precludes an identification as a binary/blend containing a WCL
star. Recent high resolution spectra obtained by us (Negueruela \& Clark 2004, in prep) confirm the identification
of C\,{\sc ii} emission, while also  revealing the presence of weak He\,{\sc i} P Cygni emission profiles.
 Such a morphology, with a lack of  N\,{\sc iv} 7116 emission, is indicative of either a very late WN star
(WN9 or later) or an extreme  early BIa+ supergiant, the lack of a He\,{\sc ii} transition in the current data
precluding a further classification.

\subsection{The Yellow Hypergiants}

The remaining 6 spectra in Figs. 3 \& 5 are seen to be dominated
by low ionisation metallic photospheric lines, suggesting spectral
types later than B. Comprehensive classification criteria and
standard spectra for cool supergiants  are provided by Munari \&
Tomasella (\cite{munari}). Comparison to their data,  employing
the relative strengths  of the Ca\,{\sc ii} lines and adjacent
Paschen series lines  - providing the same diagnostic as the
Ca\,{\sc ii} H \& K versus the Balmer line strength in the optical
region - achieves a broad classification for B8-F8 supergiants.
This may then be further  refined by consideration of the presence
and strength of the N\,{\sc i} transitions between 8650 to
8750{\AA} for A supergiants and the appearance of Fe\,{\sc i} and
Si\,{\sc i} lines for the F supergiants in our sample. When
combined with the (lower resolution) observations of Cenarro et
al. (\cite{cenarro}) we estimate a classification accuracy of
$\pm$2 spectral sub-types for our current observations, from which
we identify  two A  and four F stars (Tables 1 \& 2).

The narrow line widths, forming a smooth progression from the OB
supergiants, together with the pronounced N\,{\sc i} absorption
features (c.f. Figs. 17-25 of Munari \& Tomasella \cite{munari})
clearly justifies at least a supergiant classification for the 6
stars in question.  However, determination of  the bolometric
luminosity of  A-G stars may be made directly via the strength of
the  O\,{\sc i} 7774{\AA} blend;  a relationship first identified
by Merrill (\cite{merrill}). We make use of the most recent
calibration of this  relationship by  Arellano Ferro et al.
(\cite{arellano}; their Eqn. 2); which is calibrated from $M_{\rm
V}$=+0.35 to $-$9.5~mag. The results are presented in Table 2,
with four stars (W4, 12a, 16a \& 265) apparently demonstrating
absolute visual magnitudes in excess of the current calibration
for the relationship. Consequently, we simply limit ourselves to
the conservative statement that these stars have M$_V \sim -$9.5
(log(L$_{\ast}$/L$_{\odot}$)$\sim$5.7); placing them at the
empirical Humphreys-Davidson limit for cool hypergiants (under the
assumption of negligible bolometric correction).  We further note
that both the presence of chemically processed material at the
stellar surface and/or  infilling of the blend due to emission
from a stellar wind will  only cause a systematic underestimate of
the true luminosity; such an effect is apparently observed for the
YHG \object{$\rho$ Cas}. Therefore, we conclude the six A-F stars
in Wd~1 have luminosities that are {\em at least} directly
comparable to those of the  field populations of YHGs found in
both the galaxy and the Magellanic Clouds.

However, as described by de Jager (\cite{dejager}) a high
luminosity ($M_{\rm V} > $ -7) is not sufficient to designate a
star as a {\em bona fide} YHG. Rather, he defines  the
`Keenan-Smolinski' criteria for the spectroscopic classification
of hypergiants; (i) the presence of one or more broad components
of H$\alpha$ and (ii) absorption lines that are significantly
broader than those of Ia stars of similar spectral type and
luminosity. Both criteria are designed to identify those stars
with an enhanced mass loss rate, which de Jager (\cite{dejager})
introduces as the defining criteria of YHGs. Comparison of our
spectra to those of {\em bona fide} YHGs demonstrates that their
line widths are indeed consistent with such a classification (H.
Nieuwenhuijzen priv. comm. 2003).

Between 2001-02, H$\alpha$ has only been observed in emission in
W16a, although West87 report emission in both W16a and W265,
suggesting that it may be a transient phenomenon. Indeed,
transitions from absorption to emission have been  observed in
\object{$\rho$ Cas} (Lobel et al. \cite{lobel}), suggesting  that
the lack of H$\alpha$ emission in the majority of the stars
considered here may not disqualify them from a hypergiant
classification. Moreover, we note that radio observations of the
cluster (Dougherty et al.  \cite{dougherty}) reveal all six to
have radio counterparts which  we suggest to be the result  of the
ambient UV radiation field provided by the hot star population of
the cluster ionising their  stellar winds (given that the stars
themselves are likely too cool to ionise their own winds). We
interpret this observation as an alternative diagnostic for mass
loss from the stars and hence propose a hypergiant classification,
noting that all six objects lie on the border or within the
`yellow void', a region of the H-R diagram occupied by known YHGs
(e.g. Fig. 1 of de Jager \cite{dejager}).

\begin{table}
\begin{center}
\caption{Equivalent widths for the O\,{\sc i} $\sim$7774{\AA} blend for the six yellow hypergiants in Wd~1,
with absolute visual magnitude and resultant luminosity determined directly from the relationship of Arellano Ferro
et al. (\cite{arellano};  formal error of $\pm$0.4~mag.).}
\begin{tabular}{cccc}
\hline
YHG & Spec. & EW(O\,{\sc i}) & M$_v$  \\
    & Type  &    ({\AA}  &  \\
\hline
4   & F2Ia$^+$ & 3.0 & -10.1  \\
8   & F5Ia$^+$ & 2.1 & -8.7 \\
12a  & A5Ia$^+$ & 2.8 & -9.9  \\
16a  & A2Ia$^+$ & 2.7 & -9.8 \\
32  & F5Ia$^+$ & 2.3 & -9.1 \\
265 & F5Ia$^+$ & 2.6 & -9.7 \\
\hline
\end{tabular}
\end{center}
\end{table}

\subsection{The Red Supergiants; W20, 26 \& 237}

Finally we turn to the three M supergiants within Wd~1. Unfortunately, our current low resolution data do not
allow us
to improve on the classification of West87 for these objects. However, unlike West87, who concludes
that W237 is seen in chance projection against Wd1, we suggest that it is in fact a bona fide member of the cluster. We
base this  conclusion not only  on the comparable  optical-mid-IR fluxes of the three M stars, but, more
compellingly, the fact that  all
three are strong, spatially resolved radio sources (Dougherty et al. \cite{dougherty}). To the best of our knowledge
no other RSGs have similar radio properties (e.g. Clark et al. \cite{clark98}); hence we consider it likely that this
phenomenon results from membership of Wd~1, in a manner identical to that described above for the YHGs
(Sect. 3.4).

The spectrum of W26 also demonstrates a strong double peaked
emission line at 6564/82{\AA} (EW=$-$74{\AA}) that we attribute to
a blend of H$\alpha$ and [N\,{\sc ii}] 6583.5{\AA} emission,
noting that West87 also identifies H$\alpha$ in the spectrum of
W26 {\em and} W20. Two narrow emission lines at 9062 \& 9525{\AA}
are also observed in the spectrum of W26. Owing to the difficulty
of defining  continua in these regions of the spectrum we have
been unable to determine an EW for either line, but tentatively
identify  them as [S\,{\sc iii}] 9069.4 \& 9532.5{\AA}. W26 is
known to be identified with extended radio and mid-IR emission
(Clark et al. \cite{clark98}) with an apparent bow shock
morphology and we consider it likely that the emission arises in
this nebular material. Identical emission lines are also observed
towards  two stars $\sim$1.5~arcmin to the West (with colours
consistent with membership of Wd~1), which are coincident with a
an extended region of radio emission (Dougherty et al.
\cite{dougherty}). Thus we suggest  a `nebular' rather than
`stellar' origin  for the emission lines, although we note that
the nebular material associated with W26 likely results from the
interaction of its stellar wind with the intracluster medium/wind.

\section{Photometric results}

\begin{figure*}
\vspace*{10cm}
\includegraphics{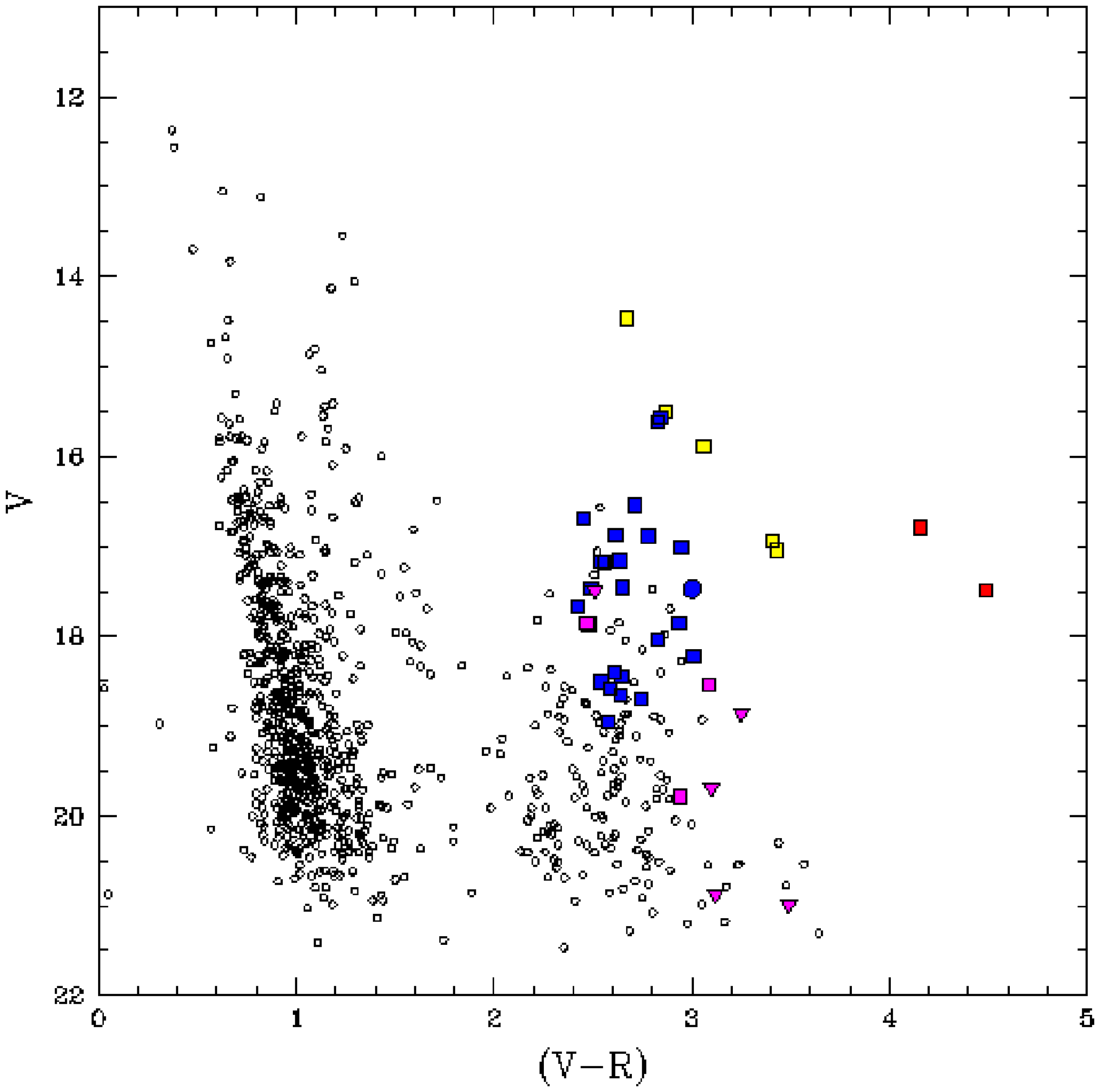}
\includegraphics{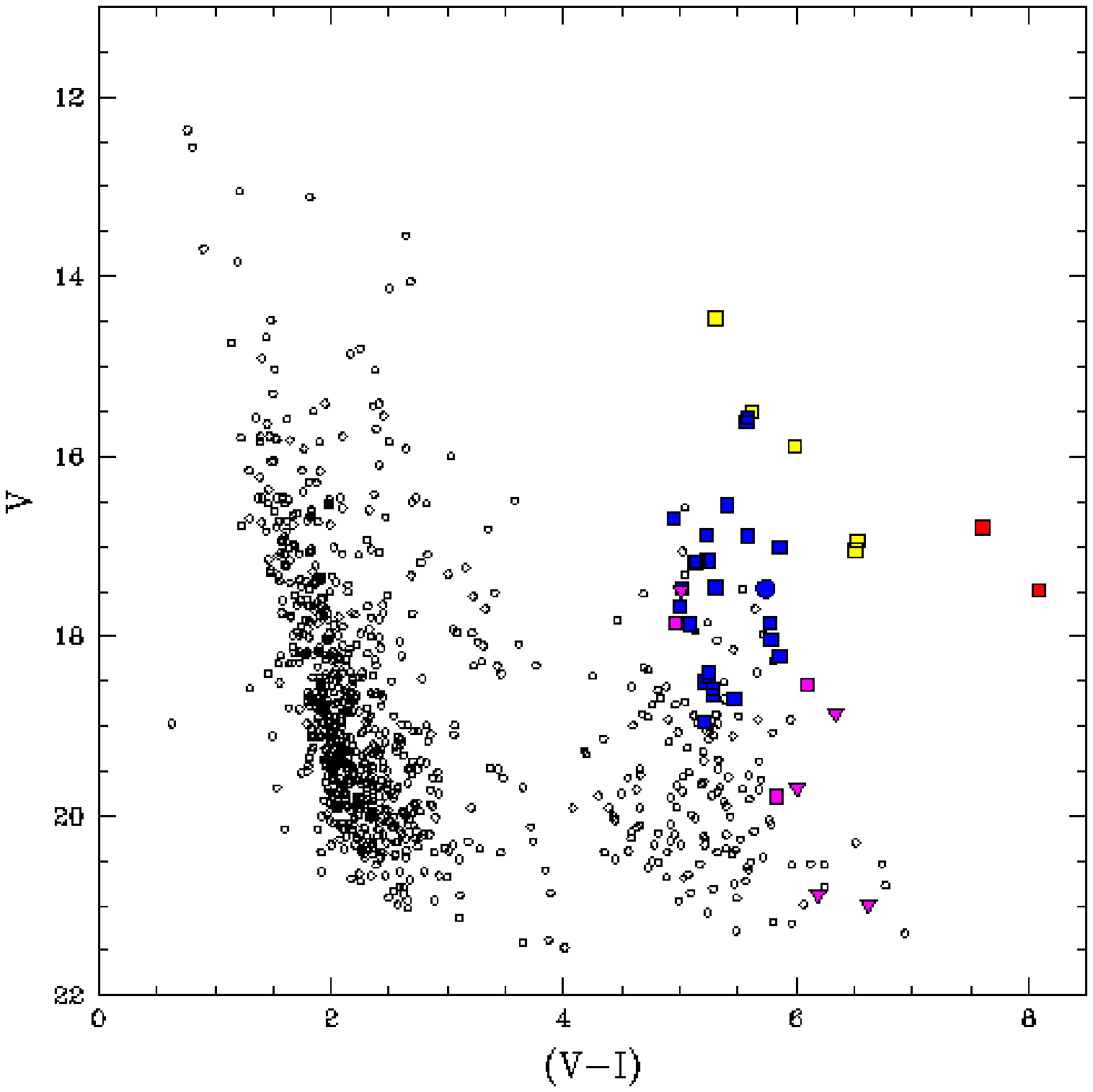}
\caption{Plot of $V$ band magnitudes versus the $(V-R)$ and $ (V-I)$ colour indices for
the 5.5$\times$5.5~arcmin field centred on
\object{Wd~1}. Objects with spectroscopic classifications are indicated with
filled symbols with  red, yellow and blue squares for RSGs, YHGs and OB SGs respectively
(the sgB[e] star \object{Wd~9} is indicated by a circle). WRs are indicated by magenta symbols,
 with squares WC and  triangles WN stars; the 2 candidate WNL stars W5
\& 44 are included in this category.
 Note that photometry is lacking for a number of cluster members such as
the LBV  \object{W243},  YHG \object{W32}, the RSG \object{W20}
and 6 WRs,
 while the two anomalously luminous OB SGs are the extreme B5 Ia$^+$ objects
W7 \& 33.}
\end{figure*}

In Fig. 9 we present  V vs. (V$-$R) and (V$-$I) colour magnitude
plots for the 1044 stars within the 5.5$\times$5.5 arcmin field
containing \object{Wd~1}, with photometric errors derived from
{\sc daophot} presented for BVRI bands  in Fig. 10. Several
distinct populations of stars are  clearly delineated in both
colour magnitude plots. Through comparison to the similar plot for
the highly reddened cluster
 \object{Pismis 23}  (Piatti \& Claria \cite{piatti02})  we identify  a long blue tilted Main Sequence (MS) for field
stars from V=12-21~mag.  and (V$-$R)=0.6-1.4 and  a vertical plume
centred on (V$-$R)=1.2, which we identify with a corresponding
population of Horizontal Branch stars. A second, more sparsely
populated MS  is apparent at (V$-$R)$\sim$1.2-1.6 - suggesting the
presence of  an intervening absorber\footnote{The photometric data
of West87 - which were obtained for a greater field of view -
reveal a much richer second MS, suggesting the obscuring cloud is
of limited spatial extent, apparently covering a region not much
larger than the cluster.}. Finally, we find a large population of
highly reddened stars with (V$-$R)$\geq$2 ((V$-$I)$\geq$4) that we
propose as cluster members.

Photometry for those cluster members with spectroscopic identifications is provided in Table 1.
 Unfortunately photometry of 12 spectroscopically identified cluster members was unavailable
due to severe blending  or the star falling in the gap between the
two detectors. The LBV W243 is  one particularly regrettable
example  of the latter failing, as it had not been  recognised as
a  variable star when the observations were made.  In total we
have currently  identified  201 probable cluster members via a
combination of spectroscopic and photometric observations.

\subsection{The Reddening \& Distance to Wd~1}

As  Piatti  et al. (\cite{piatti98}), we find a scatter in the
optical colours of cluster members that clearly exceeds the
photometric errors, and implies significant differential
extinction across the cluster. However, we note one essential
difference between Piatti et al. and our results - we find no
evidence for a Main Sequence in our current photometric data set.
We may use our spectroscopically identified OB stars to study the
reddening to Wd~1 -  given that the optical colours of all early
OB supergiants are rather similar - finding no systematic
gradients in reddening across  the cluster.  In Fig. 11 we plot
the (V$-$R) vrs. (V$-$I) colours of the early and mid-OB
supergiants to  demonstrate the extent of the differential
reddening. We find E(V-R) and E(V-I) to range from $\sim$2.6-3.2
and $\sim$5.3-6.2 with median values of 2.76 and 5.65
respectively\footnote{Note that since the differential reddening
is a real physical effect we present a range of values rather than
a formal error derived from the standard deviation of the datatset
and we choose to use the median rather than the mean value of
colour excess since we find that latter to be skewed by a few
stars with particularly high reddenings.}.   We also plot  least
square fits to the data, allowing us  to derive the reddening
vector for  the cluster. The resultant reddening vector does {\em
not} pass through the  locus for unreddened OB supergiants,
indicating a {\em non standard extinction law towards Wd~1}; a
fact we return to below.

Given that a spectroscopic discriminant exists for the intrinsic luminosity of the YHGs, we may use these stars to
determine V-$M_{\rm V}$ to the cluster. Again, due to differential reddening we find this to range from 24-26.5~mag. -
the
latter value for the isolated star W265 to the North of the cluster (noting a
formal error of 0.4~mag. due to the uncertainties in the calibration of the $M_{\rm V}$ -- EW$_{OI 7774}$
relationship).
The mean and median values of V-$M_{\rm V}$ are both $\sim$25.3~mag., which we
adopt for the remainder of the paper.
Piatti et al. (\cite{piatti98}) derive a  value of  V$-M_{\rm V}$=23.8$\pm$0.3
- at the lower range of values we find - via isochrone fitting to
the sequence of OB supergiants within  Wd~1, which they mistook for a {\em bona fide} MS.
 They noted the difficulty of such a task given that their erroneous  MS is both $\sim$vertical
and  significantly broadened in Wd~1 due to differential reddening. Their  vertical placement of the 4~Myr
isochrone was therefore accomplished via reference to the presence of the as then  currently identified supergiants,
for  which Piatti et al. (\cite{piatti98}) significantly underestimated
their absolute visual magnitudes, leading to
a corresponding underestimate of  V$-M_{\rm V}$.

Two other methods employed by Piatti et al. (\cite{piatti98}) to infer V$-M_{\rm V}$  - and hence the reddening and
distance - for  Wd~1   were comparison of their
photometry to the values of $M_{\rm V}$ quoted by West87 for the brightest cluster members and a comparison to
template
clusters. The  former suffered from the significant underestimation - by $>$1~mag. on average - of the relevant
M$_{\rm V}$s by West87, while the
latter invokes a  mistaken comparison to the 10-15Myr template open clusters \object{NGC 457} and \object{NGC
884}; both cases leading to underestimates of  V$-M_{\rm V}$.

Adopting  V$-M_{\rm V} \sim$25.3mag. to Wd~1  clearly demonstrates
why no MS objects have yet been  spectroscopically identified. An
O7 V star (Sect. 5.1) with an absolute magnitude of V=$-$4.7 would
have an apparent  magnitude  of  V$\sim$20.6~mag, at the limits of
detectability for our current photometry and beyond the reach of
our current spectroscopy. Consequently, we are led to the
conclusion {\em that the majority of cluster members currently
identified are post-Main Sequence objects.}

Finally, we may attempt to  determine the distance to the cluster
by individually dereddening the YHGs, for which we have a
spectroscopic luminosity discriminant (which is lacking for the OB
supergiants). This yields a mean  distance of  $\sim$5.5~kpc and
a mean A$_v \sim$11.6~mag. We regard this distance estimate  as an
{\em upper} limit for three reasons. Firstly, increasing the
distance would  yield correspondingly larger  luminosities for
both  the YHGs and RSGs, placing them above the empirical
Humphreys-Davidson (HD) limit for cool stars. Indeed such a large
distance - possibly biased by an overestimate of the luminosity of
the isolated YHG W265 - already places W16 $\sim$0.5 magnitudes
above the HD limit. Secondly, the distance determination  assumes
a standard reddening law which we know to be incorrect. From the
median E(V-I)$\sim$5.65 we find A$_v$=11.0, via the ratio
E(V-I)=1.6E(B-V) given by e.g. Fitzpatrick (\cite{fitz}) and
A$_v$=3.1E(B-V); appropriate for a standard reddening law.
However, from the OB supergiants we measure a median
E(B-V)$\sim$4.35, resulting in an A$_v$=13.6. Therefore we
conclude that the above methodology results in an underestimate of
the true reddening, leading in turn to an overestimate of the
distance to Wd~1. The final reason to favour a distance  of
$\leq$5.5~kpc is that anything greater would place  Wd~1 within
4~kpc of the Galactic Centre, a region underpopulated by H\,{\sc
ii} regions from which it might have formed (e.g. Fig. 5 of
Russeil \cite{russeil}).

A lower limit to the distance may be determined from the lack of radio detections of the Wolf Rayet population of
Wd 1 (Dougherty et al. in prep.). Assuming a mass loss rate of $\sim$10$^{-5}$M$_{\odot}$yr$^{-1}$ and a
wind velocity of 3000kms$^{-1}$ for the WRs, we would have obtained 3$\sigma$ detections even at a distance of 2kpc
(Dougherty priv. comm. 2004). Consequently we  adopt 2kpc as a lower limit to the distance of Wd~1
\footnote{We note that even this conservative estimate for the minimum distance to Wd~1 is significantly in excess of
the
1.1$\pm$0.4~kpc finally adopted by  Piatti et al. (\cite{piatti98}). Such a low distance estimate resulted from the
errors in determining V$-M_{\rm V}$ described above and also the value adopted for the E(B-V)/E(V-I) ratio.
Specifically,
Piatti et al. (\cite{piatti98}) employ the calibration of Dean et al. (\cite{dean}), while we have adopted the more
recent calibrations of Rieke \& Lebofsky (\cite{rieke}) and Savage \& Mathis
(\cite{savage}), which have
subsequently been confirmed by other studies (e.g. Cardelli et al. \cite{cardelli} and Fitzpatrick
\cite{fitz}).}.

\subsection{The structure of Wd~1; spatial extent and mass segregation}

In Fig. 12 we present a plot of the positions of potential cluster
members based on the colour cut  offs above.  Previous analysis by
Piatti et al. (\cite{piatti98}) had suggested that Wd~1 has a
core+halo structure with  radii  of $\sim$36'' and $\sim$72''
respectively, with a radial light profile that diverges from that
expected for a King profile.  From Fig. 12 we find  that $\sim$50\% of
the cluster members  are found within a $\sim$50 arcsec diameter
`core',  located approximately at R.A.=16 47 04 $\delta$=$-$44 09 00
(J2000), corresponding to $\leq$1.2pc for a distance of
$\leq$5.5~kpc. An apparent deviation from  spherical symmetry is
present, with an overabundance of stars to the South West - however
due to the requirement to site the cluster away from the gap between
the CCDs we lack information on the extent of the cluster in this
direction. To the North East we find a few candidate cluster members
out to radii of $\geq$3arcmin (as measured from our  cluster 'core').

However, these values were determined solely from analysis of the
distribution of the cluster supergiants, thus excluding both the
more massive and evolved WR  component, and also the currently
undetectable Main Sequence (M$ \leq$30M$_{\odot}$; Sect
5.1). Consequently, we  refrain from drawing firm conclusions on the
extent and cluster density profile.  We do  not expect the positions
of the currently observed post-MS population to accurately represent
the underlying stellar population, as many young clusters such as
R136/NGC 2070 show evidence of strong  mass segregation (e.g. Meylan
\cite{meylan}, Schweizer \cite{schweizer}).  Indeed, by analogy with
other young massive clusters we might  expect the density profile of
Wd 1 to be described by a EFF (Elson  et al. \cite{elson}) King-like
profile  (Schweizer \cite{schweizer}; Larsen \cite{larsen}); this will
be  addressed in a  future paper in which we analyse VLT (NAOS CONICA
\& FORS1) imaging data.

\begin{figure}
\resizebox{\hsize}{!}{\includegraphics[angle=0]{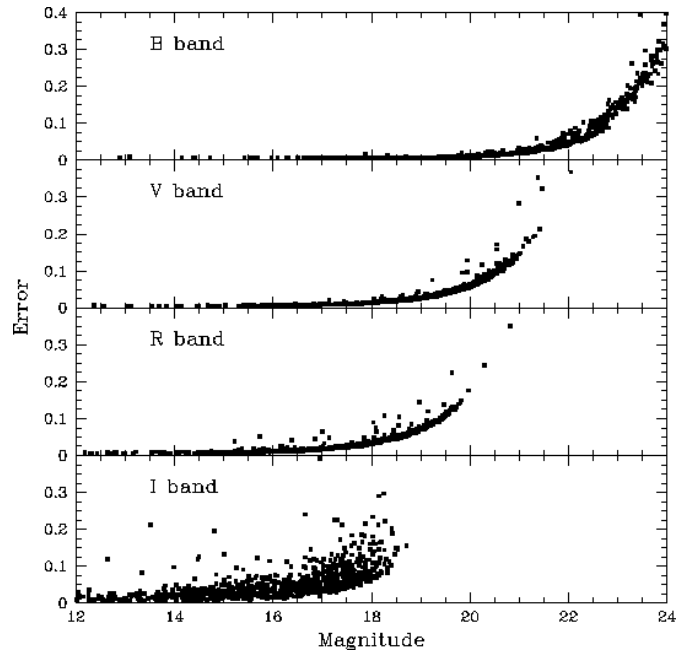}}
\caption{Magnitude errors from {\sc daophot} for the cluster
field.}
\end{figure}

\begin{figure}
\resizebox{\hsize}{!}{\includegraphics[angle=-90]{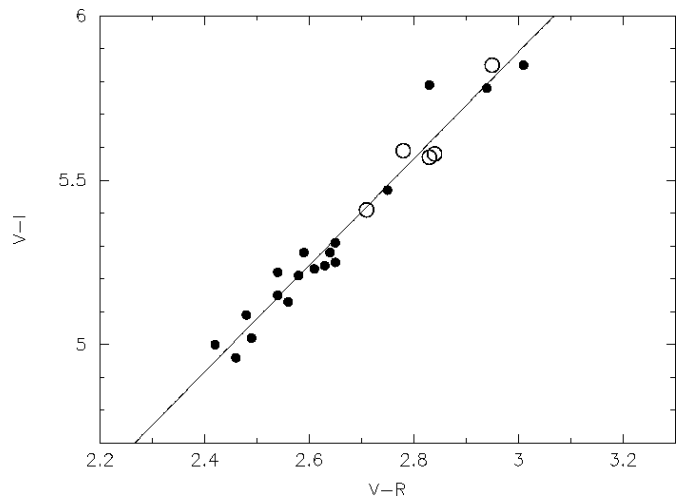}}
\caption{Plot of the (V-R) and (V-I) colours for the O9.5  to B0.5
(filled symbols)
 and B3 to B5  supergiants (open), demonstrating the extent of the differential reddening across the cluster.
Best fits to the reddening vector of the early and early+mid B supergiant datasets are also plotted and
are found to be  almost indistinguishable.}
\end{figure}

\begin{figure}
\resizebox{\hsize}{!}{\includegraphics[angle=-90]{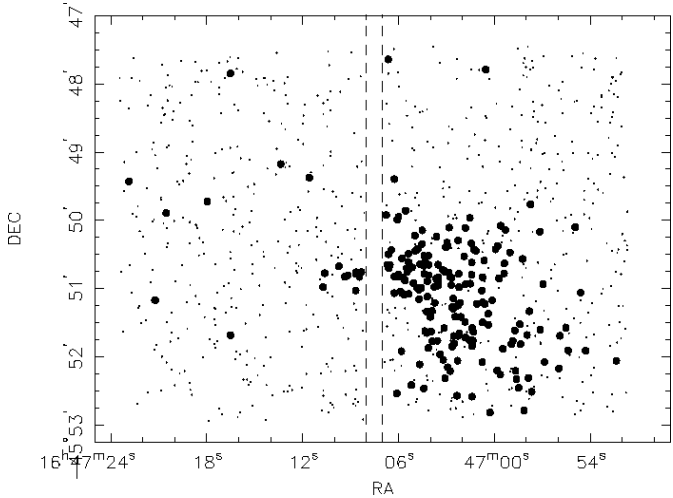}}
\caption{Plot of the positions of the photometrically selected cluster members (solid circles) and field
stars (dots). The position of the gap between the two CCDs is indicated by the dashed lines.}
\end{figure}

\section{Discussion}

Spectroscopic and photometric analysis of the upper reaches of Wd~1
reveal a  unprecedented rich  population of massive  post-MS
objects. Prior to this study, the three Galactic Centre clusters had
hosted the richest population of such stars in the galaxy. In Table 3
we summarise the {\em  spectroscopic} findings for Wd1 and the
populations of the Quintuplet, Arches and Galactic Center clusters.
Clearly, Wd~1 rivals, if not exceeds the stellar content of each of
these clusters. While OB supergiants and WRs  are found in all four
clusters, a rich  population of cool transitional objects is unique to
Wd1; indeed,  the population of  YHGs in Wd1 is equal to  that
found for  the  Galaxy or the  LMC. Given the apparent brevity of this
evolutionary phase, inferred from  the rarity of these stars, such  a
population suggests  that we are still incomplete for both their
likely progenitors - OB supergiants - and descendants - WR stars.

Comparison to the current photometric dataset supports this
conclusion. Assuming an average $M_{\rm V} \leq -7$ and  V$-M_{\rm V}
\sim$25.3~mag. yields a total  of 23 candidate OB supergiants with
V$\leq$18~mag. (excluding known cool  transitional objects); we have
spectroscopically confirmed the identity of only  $\sim$60\% of this
sample. Moreover, we note that a further 12 of the spectroscopically
confirmed OB supergiants either lack  photometry due to blending in
the crowded `core' regions  or are fainter than V$\sim$18 - the
faintest being W15,  with V$\sim$19~mag. Between V=18-19~mag. we have
spectroscopy of only 10 of 40 potential OB supergiants, although  we
note that an accurate derivation of a limiting magnitude for
supergiant candidates will  necessarily be a function of the
significant differential  reddening across Wd~1.

Moreover, we expect that a proportion of the stars found between
V=18-20~mag will likely be post-MS but of lower luminosity classes -  Ib
or II - and hence intrinsically fainter than the Ia stars identified to
date. Given the limitations of the current spectroscopic data set, we
cannot yet test this hypothesis,  but we suggest that the the range in V
band magnitudes observed for OB stars with apparently identical colours
(and hence extinctions) supports a range of instrinic luminosities for
these stars.

More extreme results are found for the WRs. Only two stars are
found to have V$<$18~mag, six from a total of 159  cluster members
have V$>$18mag. and six are photometric non-detections. Moreover,
with the exception of the WNE star W72, all are  either WNL or WCL
- if present WNE or WCE stars might be expected to be $\geq$2~mag.
fainter in the optical - well  below our current detection
threshold.

We therefore conclude that the current spectroscopically
identified post-MS population for Wd~1, while complete  for the
cool transitional objects is significantly incomplete for cluster
OB  supergiants and WRs, ignoring the effects of possible binarity
and the incompleteness of the current photometric  dataset due to
crowding. Indeed,  under the assumption of a Main  Sequence
turnoff at $\sim$O7 V  (Sect. 5.1; currently undetectable at
V$\sim$20.6~mag), a large majority of  the currently  identifiable
cluster population will consist solely of post-MS objects.

\begin{table}
\begin{center}
\caption{Comparison of the stellar contents of Wd~1 and the three
Galactic Centre clusters (Figer \cite{figer04} and refs. therein).
For brevity we define the `Early Transitional' category to
encompass sgB[e], LBV and for Wd~1 the highly luminous mid-B
supergiants W7, 33 and 42. The stellar content presented for Wd~1
is restricted to those objects for which we have spectral
classifications; hence while we are likely complete for the cool
transitional  objects the photometric data suggest we are severely
incomplete for both OB supergiants and WRs. We have explictly
excluded the WNL/B Ia+ star W5 and the WNE star possibly present
as a blend in the  W14a spectrum from the count of WR stars in
Wd~1. Note that the WC content of the Quintuplet includes the five
Quintuplet Proper Members, while for the Galactic Center cluster
Genzel et al. (\cite{genzel}) propose 6 stars with Of/LBV
classifications which we include as early transitional objects -
hence the apparent absence of any OBIa stars for this cluster.}
\begin{tabular}{ccccccc}
\hline
 & OBIa & Early  & YHG & RSG & WN & WC \\
 &      & Trans. & & & & \\
\hline
{\bf Wd~1}  & $\geq$25  & $\geq$5 & 6 & 3 & $\geq$7 & $\geq$6 \\
Quintuplet & 14 & 2 & 0 & 1 & 5 & 11 \\
Arches & 20 & 0 & 0 & 0 & $\geq$6 & 0 \\
Center & - & 6 & 0 & 2 & $\geq$10 & $\geq$10 \\
\hline
\end{tabular}
\end{center}
\end{table}

\begin{table}
\begin{center}
\caption{Comparison of inferred properties of Wd~1 to other massive young clusters (after Table 5 of
Figer et al. \cite{figer}). $M1$ is the mass in observed stars and $M2$ is the total
mass assuming  a Salpeter IMF between 1-120M$_{\odot}$.}
\begin{tabular}{lcccc}
\hline
 & log $M1$ & log $M2$  & Radius  & Age   \\
Cluster & (M$_{\odot}$) & (M$_{\odot}$) & (pc) & (Myr) \\
\hline
{\bf Wd~1}  & 3.8 & 4.75 & 0.6 & 3.5-5  \\
Quintuplet & 3.0 & 3.8 & 1.0  & 3-6 \\
Arches & 3.7 & 4.3 & 0.19 &  2-3 \\
Center & 3.0 & 4.0 & 0.23  & 3-7 \\
NGC 3603 & 3.1 & 3.7 & 0.23  & 2.5 \\
R136  & 3.4 & 4.5 & 1.6  & $< 1-2$ \\
\hline
\end{tabular}
\end{center}
\end{table}

\begin{figure}
\resizebox{\hsize}{!}{\includegraphics[angle=-90]{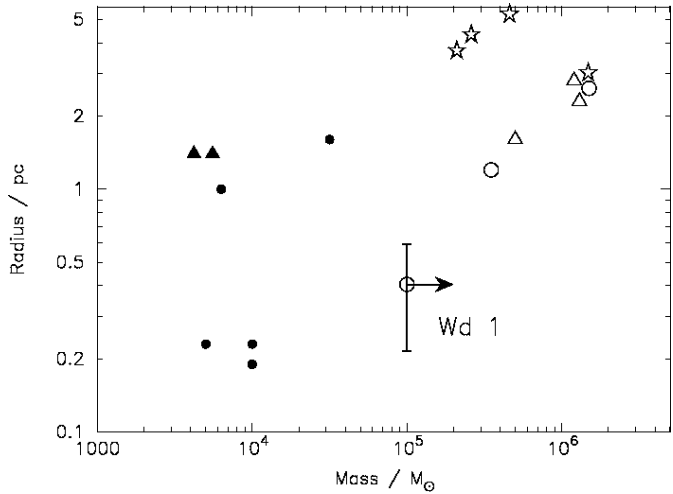}}
\caption{Plot of the mass and radius of Wd~1, selected young massive clusters in the Galaxy and LMC and SSCs for which
dynamical estimates of their mass are available. Data from Figer et al. (\cite{figer}; filled circles),
Slesnick et al. (\cite{slesnick}; filled triangles), Smith \& Gallagher (\cite{sg}; open triangles), McCrady et al.
(\cite{mccrady}; open stars) and Larsen et al. (\cite{larsen}; open circles).}
\end{figure}

\subsection{Progenitor masses for the cluster members}

Unfortunately, the current data do not permit the accurate
construction of an HR diagram for Wd~1. This is  due to the
difficulty in accurate spectral classification of the early OB
supergiants from our present  spectroscopic dataset.  Currently we
are  unable to constrain the spectral types of the stars
comprising this  population to better than  O9.5 to B0.5 Ia - or a
range in temperature of 6000K, corresponding to an uncertainty  in
bolometric correction of $>$0.6~mag. A similar problem is also
present for the population of WRs and is further compounded by the
lack of photometry for the majority of these stars. We further
suspect that many of the  WRs with detections are likely to be
binary systems, a problem - compounded by blending - that also
likely afflicts our sample of OB supergiants. Consequently,
inferring stellar masses and a cluster age from the current
dataset is difficult, particularly due to the lack of a detectable
Main Sequence. Nevertheless, assuming co-evality - justified in
Sect. 5.2  -  we may make progress in determining likely masses
for many of the cluster objects.

Trivially, the population of WRs are the initially most  massive
objects currently identified in Wd~1. From a  study of 12 galactic
clusters Massey et al. (\cite{massey01}) report  that both  WNL and WC
stars are found in clusters with  the highest turnoff mass
($\geq$50~M$_{\odot}$), although only one WCL star is reported in
their study. By  contrast, WNE stars are identified in clusters with a
wide range of turnoff masses, with  a lower limit  of only
$\sim$20M$_{\odot}$.  Following the results of Massey et
al. (\cite{massey01}) the currently  identified WR population of Wd~1
-  predominantly consisting of WNL \&   WCL stars - are likely
descended from similarly high mass progenitors.

Of the transitional stars, comparison of the mean luminosity
inferred for the YHGs  (log($L/L_{\odot}$)$\sim$5.7) to
theoretical predictions (e.g. Meynet \& Maeder \cite{meynet})
suggests  progenitor masses  of $\sim$40M$_{\odot}$. This is
consistent with the observation that galactic YHGs occupy a
rather narrow range of luminosities -
log($L/L_{\odot}$)$\sim$5.6-5.8 (e.g. Smith et al.  \cite{smith})
- and hence progenitor masses. In this work Smith et al. suggest a
possible evolutionary sequence for such stars  of:\newline

 O MS$\rightarrow$BSG$\rightarrow$RSG$\rightarrow$YHG$\rightarrow$WN9-11$\rightarrow$WR \newline

If this scenario is correct, we might also expect the remaining
cool and hot transitional objects to share a similar mass and
luminosity with the YHGs.

One possible objection to this hypothesis might be the apparent
presence of only a single LBV in Wd~1, a phase  which is known to
occur for stars in excess of log($L/L_{\odot}$)$\sim$5.4. However,
the harsh UV rich  environment of Wd~1 is likely to be inimical to
the long term survival of LBV ejecta, such  that only long term
monitoring may reveal further examples.  Obvious candidates
include the extreme B supergiants W7, 33 \& 42a, the WNL stars W5
\& 44 and BSGs W23a \& 43a,  which apparently show evidence for
H$\alpha$ variability between the 2 epochs of our observations.

Moreover, Smith et al. (\cite{smith})  identify an apparent lack
of known LBVs  in the luminosity range occupied  by the YHGs and
suggest a novel explanation  for this - that the effect  of the
bi-stability jump on the mass  loss rate of post RSG stars
evolving bluewards into the S Doradus instability  strip is
sufficient to generate  a pseudo-photosphere, effectively
suppressing the LBV phase for  stars of $\sim$40M$_{\odot}$. Thus,
the apparent  lack of LBVs in this luminosity range may be a real
feature of the evolutionary sequence for stars of  this mass.

Finally, as described in Sect. 3.3.1 the presence of H$\alpha$
infilling or emission in the population of OB supergiants suggests
that they have a mass at the upper range -  $\geq$30M$_{\odot}$ -
of those observed by McErlean et al. (\cite{mcerlean}). Thus we
arrive at a picture of the current observed population of stars
within  Wd~1 having progenitor masses ranging from
30-40M$_{\odot}$ (the OB supergiants) through to
$\geq$40M$_{\odot}$ (the WRs), with the various cool (YHGs \&
RSGs) and hot (LBV and extreme B supergiants) transitional objects
defining a narrow range around $\sim$40M$_{\odot}$.  Assuming the
OB supergiants represent the least evolved members of Wd~1
currently identified, a Main Sequence turnoff is  likely to be
found around $\sim$30-35M$_{\odot}$, implying an $\sim$O7 V
classification. As previously described, such a population would
be below our current detection limit, thus consistent  with the
lack of any Main Sequence objects in our current photometric or
spectroscopic data.

\subsection{The age of  Wd~1}

While we may not determine an age for Wd~1 on the basis of
ioschrone fitting, we note that the presence of the WCL population
suggests a minimum age of $\sim$3.5~Myrs, while the presence of O
supergiants implies a maximum age of  $\sim$5~Myr (Meynet \&
Maeder \cite{meynet}). The high luminosity and hence mass inferred
for the YHGs suggest  an age of $\sim$4~Myrs (Clark \& Negueruela
\cite{clark02} and refs. therein), while following Figer et al.
(\cite{figer02})  the presence of relatively cool emission line B
supergiants  implies an age $\geq$4Myr. We therefore suggest a
likely age for  \object{Wd~1} in the range 3.5-5Myr, which is also
consistent with our predictions of a MS turnoff around O7 V or
later.

An age in excess of $\sim$3~Myr is supported by the lack of
diffuse radio emission coincident with Wd~1 (e.g. Dougherty et al.
\cite{dougherty}). It is expected that for clusters of this age a
cluster superwind, driven  both by the present stellar winds and
also  a possible supernova component, will have dispersed  their
natal envelopes. Examples of this phenomenon include the clusters
Danks 1\& 2 found within the G305 star forming complex (Clark \&
Porter \cite{clark04b}) and \object{HD 32228} found in the LMC
Giant H\,{\sc ii} region \object{N11} (Walborn et al.
\cite{walborn99})\footnote{We note that a second generation of
trigged star formation is associated with both the G305 and
\object{N11} complexes. Examination of both radio (Dougherty et
al. \cite{dougherty}) and mid-IR (Midcourse Space Experiment
8-25$\mu$m) data  reveals no such star formation activity
associated with Wd~1; hence we may infer a lack of significant
remnant natal material associated with Wd~1. Motivated by the
comments of the referee, we speculate that either Wd~1 formed from
the collapse of the entire natal molecular cloud (possibly
implying a higher than average star formation efficiency) or that
the superwind generated by the unprecedented population of massive
stars within Wd~1 completely dispersed the cloud remnants before a
second generation of stars could form.}.

It is difficult to imagine a scenario in which Wd~1 is not coeval, as
is strongly suggested from the currently  classified stellar
population. The presence of just a few very  massive stars would be
expected to expel the gas not yet used in star formation from a
cluster and  halt further star formation on approximately a crossing
time (e.g. Goodwin \cite{goodwin97}).  The  lack of a significant
gaseous component in Wd~1 strongly suggests that this has already
happened.  In  order for Wd~1 to be significantly non-coeval a
mechanism would have to be found that formed only  low-mass stars
before forming all of the high-mass stars in a burst.  Even if this
were the case, our  assumption of coevality for the observed high-mass
stars would still be be correct.

\subsection{The mass of Wd~1}

Currently, we have spectroscopically identified $\sim 50$ stars  with
progenitor masses $\geq 30 M_\odot$ within Wd~1,  placing an absolute
minimum mass of Wd1 at $1.5 \times 10^3 M_\odot$. Under the
assumption that all currently identified cluster members have masses
of $\geq$30M$_{\odot}$ we derive a  total mass of observed stars some
$\sim$4 times larger. Following Table 5 of  Figer et
al. (\cite{figer}), in Column 2 of Table 4  we compare  the total mass
of {\em observed} stars in  Wd~1 to  those of other young compact
clusters within the Local Group. With the exception of the Arches
and \object{R136} clusters, we  find that Wd~1 is  significantly more massive.

However, such a naked comparison takes no account of the either the
depth of the relative  observations, nor the effect that the different
ages of  the Arches and Wd1 play.  Specifically, at an age of only
$\sim$2.5Myrs the mass-luminosity relationship is monotonic for the
Arches,  resulting in the mass estimate  M1 being  complete for stars
$\geq$20~M$_{\odot}$ (e.g. Serabyn et al. \cite{serabyn}).  In
contrast we have  shown that our current observations are  only
complete for stars with initial masses $\geq$30M$_{\odot}$ in
Wd~1. Moreover, at an age of  $\sim$4Myrs, the mass-luminosity
relationship is bimodal, such that stars in excess of 40M$_{\odot}$
will either  have been lost to supernovae or will have evolved to the
WR phase. Indeed, van der Hucht (\cite{vdh}) demonstrates  that with
the exception of the latest WNLs, WRs of both flavours  have absolute
visual  magnitudes M$_v > -$5mag. - so undetectable in our current
data\footnote{We suggest that the photometric  detections  of a subset
of the cluster WRs - predominantly the WCLs - may be a result of
binarity.}. Thus, we believe that our current observations of Wd~1 are
only sensitive to stars in a narrow range of progenitor masses  -
specifically 30-40M$_{\odot}$ - and thus may not be directly
comparable to those of  the young  $\leq$3~Myr clusters such as
\object{NGC 3603}, the \object{Arches} and \object{R136} presented in
Table 4.

Following Figer et al. (\cite{figer}) we present the total mass  for
each cluster in Table 4 assuming an identical Salpeter IMF, finding
Wd~1 to be a factor of  $\sim$2 larger than the next most massive
cluster, \object{R136}. We note however, that the current
observational position on the full IMFs of massive star clusters is
unclear.  Sirianni et al. (\cite{sirianni}) find that R136 has a
Salpeter IMF that turns-over at $\sim 2$M$_\odot$, as compared to
$0.5$M$_\odot$ in the field and local star forming regions (see Kroupa
\cite{kroupa} for a review of the IMF).  However, this has been
disputed by Zinneker et al. (\cite{zinneker}) who claim a Salpeter
slope to at least $1$M$_\odot$.

Moreover, Figer  (\cite{figer04} and references therein) suggest a
significant  deviation from  a Salpeter IMF for the Arches, prompting
a downwards revison of the total mass to $\sim 10^4$M$_{\odot}$.  The
high gas temperatures (even in dense molecular clouds) due to the
strong ambient radiation field in the Galactic Centre may well be
responsible for this difference from the `standard' IMF and so the
Galactic Centre clusters may not be directly comparable to Wd~1 or
R136 which have formed in more `normal' environments.  Indeed, Larsen
(\cite{larsen}) finds that the mass-to-light ratios of SSCs are
consistent with a Kroupa (\cite{kroupa}) IMF.  Another caveat is that
these mass estimates are generally based on  the assumption of virial
equilibrium which may be significantly wrong when dealing with very
young clusters (Goodwin et al. in preparation).

In summary, it is currently unclear what the form of the IMF of
massive star clusters is, or indeed if it is universal.  Sirianni et
al. (\cite{sirianni}) do find that the IMF in fields close to R136 is
Salpeter to at least $\sim 1$M$_\odot$.  It may be that primordial and/or
dynamical mass segregation acts to make the IMF dependent on the size of
the region under consideration, and that abnormal IMFs are the result of
only considering the cores of clusters.

Nevertheless, assuming that Wd~1 has a Kroupa two-part IMF and that
 $\geq$140 of the current candidate members have masses $>30$M$_\odot$
 - conservatively defined as the number of   stars with V$<$20~mag.,
 excluding the  small WR component - then we find that its mass must
 be $>10^5$M$_\odot$. This estimates makes no account of
 incompleteness and should be regarded as a lower limit under the
 hypothesis of a standard IMF. Moreover, the additional assumption
 that our current observations only sample  the stellar population
 within a restricted ($\sim$30-40~M$_{\odot}$) mass range suggests
 that we may be significantly underestimating the total mass of Wd1.

A more reliable comparison would  be between the number count of
stars within  the 30-40M$_{\odot}$ range - corresponding to a mid
O Main Sequence star.  The Arches has $\sim$40
stars in this mass range (Figer et al. \cite{figer02}) versus
$\sim$140 for Wd~1.  If the IMFs are similar this suggests that
Wd~1 is at least $3 \times$ more massive  than the Arches.
Similarly, Massey \& Hunter (\cite{mh}) report that R136 has 34
stars more massive than 30M$_\odot$ (M$_{\rm V}<-4.5$) within 1~pc
of the cluster core, and 117 within 10~pc (the nominal size of the
NGC 2070 cluster). This makes Wd~1 significantly more massive than
R136 and at least comparable to (and probably more massive than)
the NGC 2070 cluster at the heart of the 30 Dor complex.  It
should also be noted that the currently known extent of Wd~1 is
significantly less than 10~pc.

Similar conclusions may also be drawn for \object{Cyg OB2} and 
\object{W49A}, both of which have recently
been suggested to host massive stellar populations. On the basis
of near IR photometry Knodlseder (\cite{knodlseder}) proposed that
$\sim$120 O stars (or their descendants) are found within
\object{Cyg OB2}, which, depending on the low mass cut off
adopted, yields a total mass in the range of
4-10$\times$10$^4$M$_{\odot}$. However subsequent near-IR (Comeron
et al. \cite{comeron}) and optical (Hanson \cite{hanson})
spectroscopic surveys were unable to support such a conclusion,
with Hanson concluding that many of the candidate OB stars are
likely to be field contaminants and hence that the total O star
census is unlikely to exceed 100. Moreover, at a distance of
$\sim$1.7kpc Knodlseder (\cite{knodlseder}) finds \object{Cyg OB2}
to have a radius of $\sim$30pc (with a half light radius of
$\sim$6pc), an order of magnitude greater than we currently infer
for \object{Wd~1}.

Similarly, Alves \& Homeier (\cite{alves})  propose that $\sim$100
O stars are found within the central 16$\times$16pc region of  the
giant H\,{\sc ii} region W49A. However, as with \object{Cyg OB2},
the stellar density appears to be significantly lower than that
inferred for Wd~1, with only $\sim$30 O stars found within the
most massive cluster identified (Cluster 1 of Alves \& Homeier,
with a radius of $\sim$3pc).

Therefore, despite the uncertainties in the appropriate IMF to apply, Wd~1
appears to be significantly more massive and compact than any
other local massive star forming cluster (Fig. 13). Indeed, the
only way in which to make Wd~1 comparable to, or less massive
than, other local massive clusters would be to postulate that it
has an IMF that turns-over at a very high mass (5-10M$_\odot$) so
that very little mass is hidden in low-mass stars - arguably an
equally important result.

The high mass and extremely compact nature of Wd~1 make it a clear
example of a SSC.  Wd1 is very probably more massive than the
average  old Galactic Globular Cluster ($10^5 M_\odot$), with a
mass, luminosity and density that likely fulfills the criteria
expected for proto Globular Clusters.

\subsection{Cluster dynamics and evolution}

At least half of the currently observed stellar population of Wd1 is
found within a circular region of radius $\sim$25'', corresponding
to $\leq$0.6~pc at $\leq$5.5~kpc.
Assuming  a mass  of $>10^5$M$_{\odot}$ for
\object{Wd~1} leads to a {\em minimum} density of
log$\rho$=5.0M$_{\odot}$pc$^{-3}$, compared to
log$\rho$=5.5M$_{\odot}$pc$^{-3}$ found for the
\object{Arches} cluster.  Given that the distance is likely to be 
less than 5.5~kpc and that this estimate does not account for
incompleteness (which is expected to be worst in the centre of the
cluster), we consider it likely  that \object{Wd~1} is of comparable
density to the \object{Arches}.

The collisional timescale for stars in the centre of Wd~1 is of
order $10^4$ Myr (Binney \& Tremaine \cite{binney}) assuming a
central number density of stars of a few $\times 10^6$ pc$^{-3}$
for solar-type stars. This number will be significantly reduced by
the presence of many large (in both mass and radius) stars (see
for example the simulations of Porteies-Zwart et al.
\cite{zwart}). We might expect a collision between two of the
stars in the core of Wd~1 every $10^3 - 10^4$ yrs, and so it is
not unreasonable to propose that W9 is an unrelaxed recent merger
remnant.

Recent work by Portegies-Zwart et al. (\cite{zwart}) suggest  that
under such extreme conditions runaway stellar mergers may lead to
the production of  an intermediate mass black hole. Accretion onto
such objects has been suggested by some  authors to explain the
class of Ultraluminous X-ray Sources observed in external
galaxies.  {\em If} such an object is present in Wd~1, the lack of
such a luminous (L$_x > 10^{39}$erg/s) X-ray  detection  (Clark et
al. \cite{clark98}) may indicate that either no mass donor is
associated with  it,  or that mass transfer is either transitory
or proceeds at a comparatively low rate (e.g. via direct wind fed
accretion rather than Roche Lobe overflow).

Assuming a velocity dispersion in the core of $\sim 15$ km s$^{-1}$
(expected if the cluster core is approximately virialised) would imply
that the core is $> 100$ crossing times old.  This would mean that the
core is dynamically relaxed.  It is possible that the outlying stars
visible in  Fig. 12  have been dynamically ejected from the core.
Indeed, it is quite possible that some stars may have travelled in
excess of 50pc from the cluster centre.  Whilst the core of Wd 1
should be dynamically relaxed, the outer regions are probably
dynamically young and substructure in the initial cluster may not have
been erased (e.g. Goodwin \& Whitworth \cite{goodwin}) as is thought
to be the case in the NGC 2070 cluster (Meylan \cite{meylan}).

The huge population of massive stars in Wd~1 suggests that it will
have a very significant impact on the local ISM.  We estimate that the
total number of type-II supernovae (SNII) will exceed 1500.  This implies
that Wd~1 will inject $>10^{55}$ ergs of energy from stellar winds, UV
radiation and SNII explosions on a timescale of $<40$ Myr.  The energy
flux from Wd~1 will exceed $10^{40}$ ergs s$^{-1}$ for several Myr.
The lack of radio emission and an HII region associated with Wd~1 
(Sect. 5.2) suggests  that Wd~1 already contains very little gas, presumably 
due to feedback from the massive stellar population. 
 Indeed, the impact Wd~1 has already had upon its
environment may be the reason why it has escaped identification as
such an extreme cluster for so long.

The input of energy from many hundreds of SNII within Wd~1 would
be expected to drive a significant superbubble and evacuate a large
portion of the local Galactic disc (until destroyed by the
differential rotation of the Galaxy).  Wd~1 could well eject a
significant amount of material out of the Galactic disc, driving a
fountain or creating high velocity clouds.

Wd~1 is also expected to return $>10^3$M$_\odot$ of heavy elements to
the ISM (e.g. Tsujimoto et al. \cite{tsujimoto}).  Due to the extremely high mass
of many of the stars in Wd~1 compared to the content of lower-mass
clusters, Wd~1 would be expected to return unusual abundances of heavy
elements (e.g. Goodwin \& Pagel \cite{goody}).

\section{Conclusions}

We have presented both spectroscopic and photometric observations of
the cluster Wd~1. On the basis of their  optical colours, we have
identified $\sim$200 candidate members, of which  95\% are  found to
lie within  2arcmin of the nominal cluster core, and $\sim$50\% within an inner
region of radius $\sim$25''. Adopting an upper limit to the distance  of 5.5kpc
- from consideration of both spectroscopy and photometry - therefore implies that Wd~1 is
 rather compact; the angular radii corresponding to $<$3.2 and $<$0.6pc respectively.

We have obtained
spectral classifications for 53 candidate cluster members, all of which  are found to be
post MS stars. Of these we find 14 to be WR stars of both flavours,
25 to be OB supergiants  with spectral types of approximately B0Ia,
and at least 14 transitional objects. The latter group consists of
both high (LBV, sgB[e] \& extreme BSGs) and low (YHGs and RSGs)
temperature objects. In doing this we confirm the  result of West87 in
identifying 4 YHGs within Wd1 and add a further 2 examples, leading to
a total population  of such stars that is directly comparable to that
of the Galaxy or the LMC.

 Unfortunately, we are currently unable to construct an HR diagram for Wd~1. This is due to
 the uncertainties in  the spectral classification of the OB
 supergiants, which preclude an accurate determination of their
 intrinsic luminosities, compounded by significant differential
 reddening across the cluster and the presence of a non standard
 extinction law.  Nevertheless, we may use
spectroscopic diagnostics to infer a mean luminosity for the YHGs
of log(L/L$_{\odot}$)$\sim$5.7, placing them firmly at the HD
limit and implying $\sim$40M$_{\odot}$ progenitors. Following the
recent work of Smith et al. (\cite{smith}) we  infer similar
initial masses for the other  transitional objects within Wd~1.
While no unambiguous spectroscopic luminosity diagnostics exist
for OB supergiants, following the results of  McErlean et al.
(\cite{mcerlean}) we infer progenitor masses of
$\sim$30-40M$_{\odot}$ from the presence of H$\alpha$ emission in
their spectra.

These estimates suggest that we might expect a Main Sequence
turnoff mass for Wd~1 of around 30-35M$_{\odot}$,  consistent with
a population of late O dwarfs with spectral types no earlier than
$\sim$O7 V. Adopting a median value for V$-M_{\rm V}
\sim$25.3~mag. implies that such  objects will be found at
V$\sim$20.5~mag. This is entirely consistent with the lack of MS
objects in our  spectroscopic dataset, and the apparently lack of
a MS in the photometric data. This in turn  implies that a large
majority of the $\sim$200 currently  identified cluster members
are also likely to be  post-MS objects.

Simply considering spectroscopically classified stars within Wd~1, we
find a minimum mass of  the order of  1.5$\times
10^3$M$_{\odot}$. However,   the twin assumptions  that $\sim$140
currently   identified cluster members had initial masses
$\geq$30M$_{\odot}$ and that a Kroupa (\cite{kroupa}) two-part IMF is
appropiate for Wd~1 yield  a total cluster mass of
$\sim$10$^5$M$_{\odot}$. Note that this estimate takes no account of
incompleteness in the current datasets nor that they  are likely only
sensitive to stars within the restricted $\sim$30-40M$_{\odot}$ range,
both of which will act  to  increase the mass of the
cluster. Nevertheless, Wd~1 appears significantly more massive than
any  currently identified open cluster in the Galaxy, and indeed is likely
to be more massive than the average Galactic Globular
Clusters. We conclude that Wd~1 represents the first known example of
an SSC in the  Local Group  - i.e. a factor of $\sim$1000 times closer than the next closest
SSC, \object{NGC1569-A} (Hunter et al. \cite{hunter}).

With such an extreme mass and population of massive stars Wd~1
promises to greatly advance our understanding  of the formation and
evolution of massive stars, both individually, and in the extreme
physical conditions  present in SSCs.  Given the wealth of post-MS
objects, the identification of a MS and  construction of an HR diagram
will allow accurate progenitor masses to be assigned to different
spectral  types,  permitting the refinement of current theoretical
pathways through the post-MS `zoo' and the empirical  verification of
the HD limit for a population of co-eval stars at a single
metallicity. Deep adaptive  optics observation will permit the
identification and study of sub-solar mass objects within Wd~1 - if
their  formation has not been inhibited by the population of  massive
stars.

Indeed the presence of Wd~1 within the  galaxy will provide us
with a unique insight into the physical processes occurring in the
SSCs heretofore only  identified in external starburst galaxies,
challenging both   our current theories of star formation and the
role it plays in the wider evolution of the Galaxy. For example -
assuming a
 nominal star formation efficiency of $\sim$10\% -  how
did  $\sim$10$^6$ M$_{\odot}$ of gas  collect in a region $\sim$one
pc across?  The average density of the pc$^3$ molecular cloud from
which Wd1 formed must have been in excess of $10^6$ atoms cm$^{-3}$
and hence  the star formation process in this region must have been
very different from that in local star forming regions such as Taurus
or Orion.  Recent simulations by Bonnell et al. (\cite{bonnell}) of
clustered star formation in a highly turbulent molecular cloud give
probably the closest insight into star formation in Wd 1.

\section{Acknowledgements}
We are indebted to the Peter Stetson for his help in reducing the
 photometric dataset.  We thank the referee Nolan Walborn for his insightful comments and also
 thank Rens Waters, Sean Dougherty, Alex de
 Koter and Hans Nieuwenhuijzen  for many informative discussions. We wish
 to thank ESO for their ongoing support for this research, and  also
 the support astronomers at La Silla who have been of immense help.
 IN  is a  researcher of the programme {\em Ram\'on y Cajal}, funded
 by the Spanish Ministerio de Ciencia y Tecnolog\'{\i}a and the
 University of Alicante.  This research is partially supported by the
 Spanish MCyT under grant AYA2002-00814. PAC acknowledges financial
 support from the Royal Society. SPG is a UKAFF fellow.

\appendix
\section{OB supergiant classification criteria}

Caron et al. (\cite{caron}) have recently outlined a
classification scheme for O9 to B5 Ia - V stars between $\sim$8400
to 8900{\AA}, based on both observations of standard stars and
synthetic models. We have followed this methodology in the
extension of this scheme to early O - late B supergiants, via the
construction of an extended grid of non-LTE synethetic spectra,
normalised where possible to observations of OB supergiants. We
have further extended this to include the 6500-7900{\AA} range to
encompass our full data set, although we concentrate on the I band
for our primary classification criteria.

We calculated our  grid of  OB supergiant models
with the non-LTE atmospheric code CMFGEN (Hillier \& Miller
\cite{hillier}) between T$_{eff}$=10kK and T$_{eff}$=50kK at 2000K
intervals,  accounting for
line  blanketing by H, He, C, N, O, Mg, Al,  Si, S, Fe. Our approach
follows the approach outlined for OB stars by Crowther et al.
(\cite{crowther02}), with an assumed 20kms$^{-1}$ microturbulence. A
uniform rotational broadening
of 75kms$^{-1}$ is adopted throughout.

The temperature scale for O supergiants has recently
been revised downward as a result of the incorporation of
line blanketing (Crowther et al. \cite{crowther02}; Herrero et al.
\cite{herrero}; Repolust et al. \cite{repolust}). Similarly, the scale for
early-B supergiants follows
Crowther et al. (\cite{crowther}) whilst mid and late-B stars follow
recent
unpublished work by one of us (PAC). Luminosities for individual
models follow from theoretical bolometric corrections, plus the Humphreys
\& McElroy (\cite{humphreys84}) absolute magnitude--spectral type
calibration.

We have adopted wind velocities for each subtype following Prinja et al.
(\cite{prinja}) and Lamers et al. (\cite{lamers}), and mass-loss rates
following the wind-momentum relationships from Kudritzki et al.
(\cite{kudritzki}) for B  stars\footnote{ We adopted an identical form of
the wind momentum  relationship for late B supergiants to that established
by Kudritzki et al. (\cite{kudritzki}) for mid-B supergiants.} and Puls
et al. (\cite{puls}) for O stars.

The synthetic  spectra in the range 8200 to 8900{\AA} are plotted
in Fig. 13 with, where available, standard spectra overplotted for
comparison (the spectra of standard stars are summarised in Table
5). A summary of the model parameters for each spectrum, along
with optical and near IR colours and V band bolometric correction
are presented in Table 6. Comparison of synthetic spectra to
observational data indicate an encouragingly close correspondance,
given that models have {\em not}  been tailored for the individual
standard stars.

The first requirement  of an I band classification scheme for  the current
dataset is to distinguish between Main Sequence and Supergiant spectra,
which Caron et al. (\cite{caron}) demonstrate may be accomplished via the
FWHM of both the Paschen series and He\,{\sc i} absorption features,
which are systematically narrower for the supergiants due to the
Stark effect (their Fig. 6). As  a consequence, Paschen lines shortwards
of Pa16 and the weak HeI transitions such as  8775{\AA} are
unresolvable in Main Sequence stars, while clearly present in supergiants
of similar spectral type.

Accurate spectral classification of OB stars is more problematic,
with Caron et al. (\cite{caron}) forced to rely solely on the
absolute strengths of the Paschen lines  for classification of
stars in the range O9-B5. While we too are forced to partially
adopt this approach, particularly for stars with temperatures  in
excess of 34kK (O7I), we are able to identify a number of further
temperature diagnostics in both standard and synthetic spectra. In
attempting to describe a classification scheme we adopt the
philosophy of relying on observational over synthetic data, and
the occurrence  of specific lines and particular line ratios
rather than the absolute line strengths. For a line to be
identifiable we require an EW$>$0.2{\AA} in either our synthetic
or standard spectra. In practice all lines listed satisfy both
criteria, with the exception of the Pa16+C\,{\sc
iii}$\sim$8500{\AA}  blend, which is absent in our O8Ia (34kK)
synthetic spectrum but clearly present in the standard spectra.
He\,{\sc ii} 8238{\AA} lies outside the range of the standard
spectra presented, but its occurrence is confirmed in unpublished
O star spectra of  one of us (Crowther).

Detailed criteria for spectral types O2-A0 are:

$\bullet${\bf O2 to O7 Ia (50 to 34kK)}: The spectra of the
earliest stars are dominated by the presence of He\,{\sc ii}
8238{\AA} and
 Paschen series absorption lines alone. Consequently accurate spectral
typing within this temperature range must rely solely on the
relative strengths of the Paschen series, which our synthetic
spectra indicate increases by a factor of $\sim$2 between O2 to
O7. Without adequate observations of standard stars  to normalise
our synthetic spectra to, we refrain from providing an absolute
calibration of line strengths against spectral types (noting that
fortuitously this does not affect our  conclusions for Wd~1).

$\bullet${\bf O8 to O9 Ia (34 to 32kK)}: While superficially
similar, spectral types O8 to O9 may be distinguished from hotter
O stars by the emergence of weak He\,{\sc i} absorption features,
most notably at 8775{\AA}. An additional diagnostic is the
Pa16+C\,{\sc iii} blend at $\sim$8500{\AA}, which  has a greater
depth and strength than the adjacent Pa15 feature.

$\bullet${\bf O9.5 to B0 Ia (30 to 28kK)}: O9.5 to B0 Ia spectra
are characterised by the abrupt disappearance of He\,{\sc ii}
absorption. As with O8 to O9 Ia stars, the Pa16+C\,{\sc iii}
$\sim$8500{\AA} blend exceeds the strength of Pa15.

$\bullet${\bf B0.5 to B1.5 Ia (26 to 20kK)}: Spectra consist
solely of H\,{\sc i} and He\,{\sc i} features;  as with O2 to O7
Ia stars, detailed sub classification  relies on the absolute
strength of the Paschen series (c.f. Caron et al. \cite{caron} for
a detailed analysis).

$\bullet${\bf B2 (18kK)}: The appearance of the O\,{\sc i}
8446{\AA} line with a central intensity less than the adjacent
Pa18 line acts as a diagnostic for B2 Ia stars.

$\bullet${\bf B3 (16kK)}: As B2 Ia, but marked by the emergence of
weak N\,{\sc i} absorption features in the range $\sim$8670 to
8730{\AA}.

$\bullet${\bf B5 (14kK)}: As above, but the intensity of O\,{\sc i}
8446{\AA} now exceeds that of Pa18.

$\bullet${\bf B8 to A0 (12 to 10kK)}: As B5 Ia but with the loss
of residual He\,{\sc i} absorption features, leaving a spectrum
dominated by Paschen, O\,{\sc i} and N\,{\sc i} absorption.

Finally, an additional classification criterion found in the 6500
to 7900{\AA} band is the O\,{\sc i} 7774{\AA} blend, which is
detected (EW$>$0.6{\AA}) in both standard and synthetic   spectra
of spectral types of B0.7 and later. Therefore, if available, this
additional diagnostic helps to break the degeneracy found for the
26 to 20kK (B0.5 to B1.5Ia) temperature range.

\begin{figure*}
\vspace*{10cm} \includegraphics{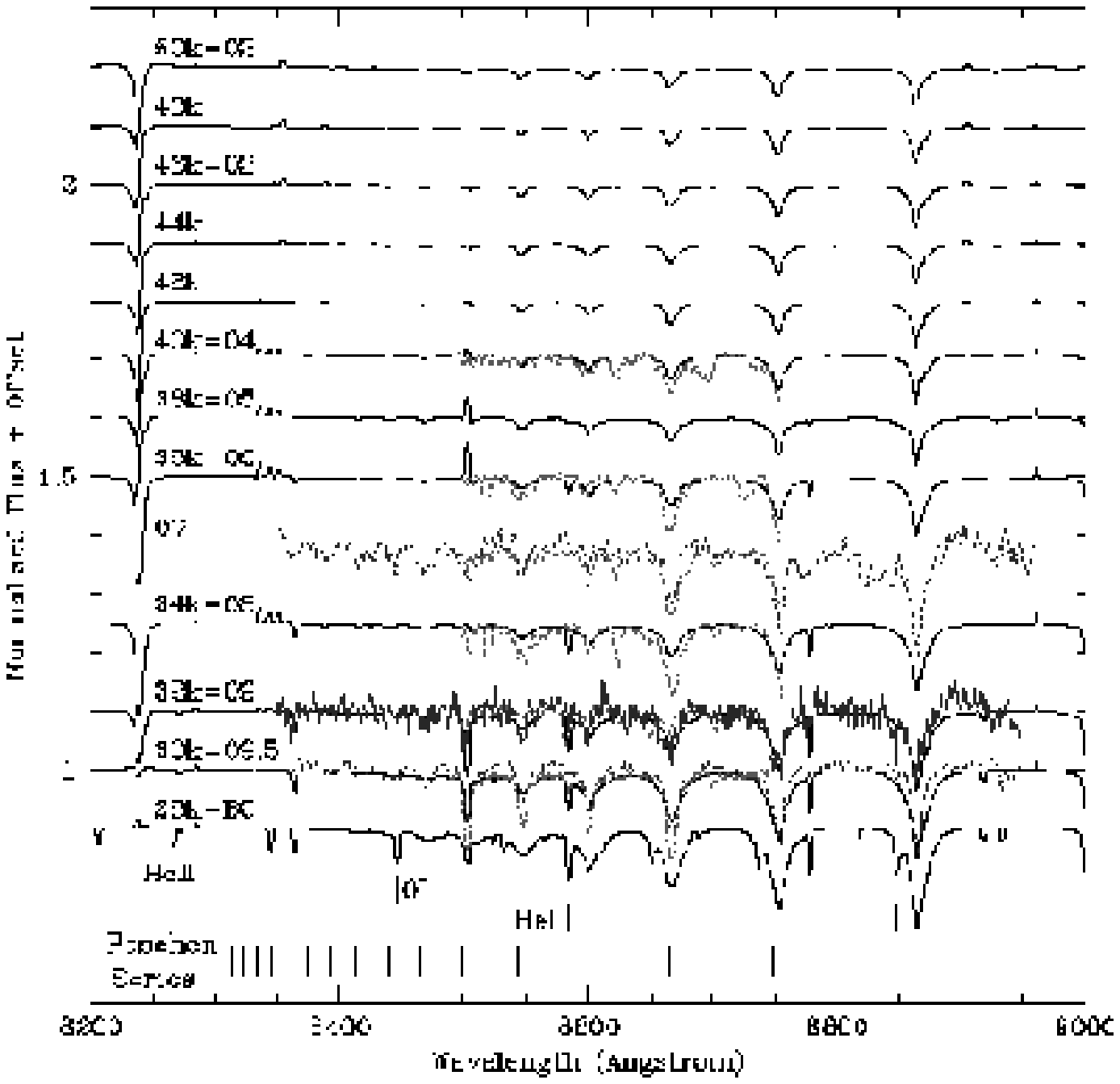} \includegraphics{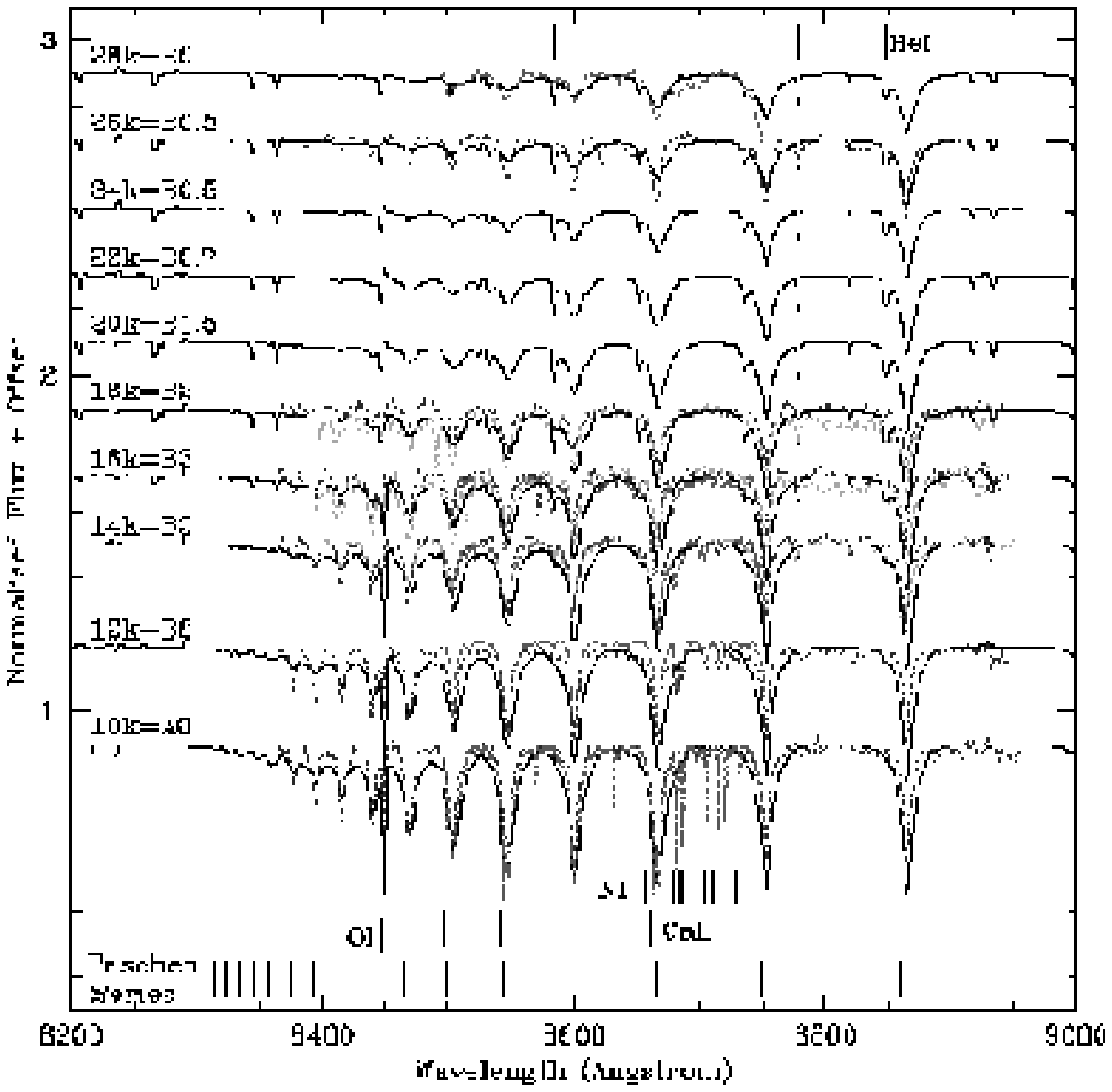} \caption{I
band (8200 to 9000~{\AA}) synthetic spectra between 10 and 50~kK
(solid black lines; parameters given in Table 3). Prominent
transitions of H\,{\sc i}, He\,{\sc i}, He\,{\sc ii},  O\,{\sc i},
N\,{\sc i} and Ca\,{\sc ii} indicated. Published OB supergiant
spectra overplotted; data from Munari \& Tomasella (\cite{munari};
red dotted lines), Cenarro et al. (\cite{cenarro}; blue dotted
lines) and Le Borgne et al. (\cite{borgne}; cyan  dotted lines);
stellar identification given in Table 2.}
\end{figure*}

\begin{table}
\begin{center}
\caption{Individual standard stars used for the construction of Figs. 4-5.
Data are from
$^1$Munari \& Tomasella (\cite{munari}), $^2$Le Borgne et al.
(\cite{borgne}),
and $^3$Cenarro et al. (\cite{cenarro}). All stars are either type Ia or
Ib supergiants, the latt
er given in itallics.}
\begin{tabular}{cc}
\hline
Spectral type & Star ID \\
\hline
O4            & \object{HD~190429}$^1$ \\
O6            & \object{HD~210839}$^1$ \\
O7            & {\em \object{HD~192639}}$^1$, \object{HD~57060}$^3$ \\
O8            & {\em \object{HD~167971}}$^1$ \\
O9            & {\em \object{HD~210809}}$^1$, {\em \object{HD~57061}}$^3$
\\
O9.5          & \object{HD~228779}$^1$,\object{HD~30614}$^1$,{\em
\object{HD~37742}}$^1$, {\em \object{HD~209975}}$^3$ \\
B0            & \object{HD~37128}$^1$ \\
B0.5          & \object{HD~194839}$^3$ \\
B2            & \object{HD~268623}$^2$, \object{HD~206165}$^3$ \\
B3            & \object{HD~198478}$^1$, \object{HD~271163}$^2$,
\object{HD~14134}$^3$ \\
B5            & {\em \object{HD~164353}}$^{1,2}$, \object{HD~13267}$^3$ \\
B8            & \object{HD~34085}$^1$, \object{HD~199478}$^3$ \\
A0            & {\em \object{HD~87737}}$^1$, \object{HD~39970}$^3$ \\
\hline
\end{tabular}
\end{center}
\end{table}

\begin{table*}
\begin{center}
\caption{Stellar properties adopted for the construction of the grid of
synthetic OB supergiant spectra presented in
Fig. 3, presented with the resultant absolute $V$  band magnitude, $(V-I)$
and $(V-K)$ colour indices, and V band bolometric correction (B.C.).}
\begin{tabular}{cccccccccc}
\hline
T$_{eff}$ & Spec  &log(L/L$_{\odot}$) & log\.{M}  & v$_{\infty}$ & $M_V$ &
(V-I) & (V-K) & B.C. \\
(kK) & Type & & (M$_{\odot}$ yr$^{-1}$) & (kms$^{-1}$) & (mag.) & (mag.) &
(mag.) & (mag.) \\
\hline
 50 & O2 & 6.24& -4.80 & 3000 & -6.5 & -0.41  &-0.84  & -4.35\\
 46 &  O3 & 6.13& -4.92& 2600 & -6.5 & -0.41  & -0.85   & -4.07\\
 40 & O4 & 5.95 & -5.18 &  2400 & -6.5 & -0.39  &  -0.83 & -3.64\\
 38 & O5 &  6.02& -5.09 & 2300 & -6.8 & -0.40  &  -0.81     & -3.49\\
 36 & O6 &  5.94 &-5.20  & 2200 & -6.8 & -0.40  &  -0.87  & -3.31\\
 34 & O8 & 6.04 & -5.08  & 2100 & -7.2 &-0.40  &-0.88  &-3.15\\
 32 & O9 & 5.97 & -5.18  & 2000 & -7.2 & -0.41  &-0.89  & -2.97\\
30 & O9.5 & 5.89 & -5.26 & 1900 & -7.1 & -0.42  & -0.88   &-2.87\\
 28 &  B0 & 5.75 & -5.8 & 1750 & -7.0 &-0.38  &-0.84  & -2.62 \\
 26 & B0.5 &  5.69 & -5.9 & 1600 & -7.0 & -0.37      &   -0.82   & -2.47
\\
 24 & B0.5 & 5.60 & -5.9 & 1200 & -7.0 &  -0.35     & -0.82      & -2.25\\
 22 & B0.7 &  5.52 & -6.0 & 1000 & -7.0 &-0.33  &-0.74  & -2.05 \\
 20 & B1.5 & 5.43 & -6.9  &  800 & -7.0 &-0.32  &-0.70  & -1.83 \\
 18 & B2 & 5.49 & -6.7  &  550 & -7.4 &-0.29  &-0.63  & -1.59\\
 16 & B3 & 5.30 & -6.9  &  400 & -7.2 &-0.22  &-0.55  & -1.30\\
 14 & B5 & 5.09 &-7.2 &  300 & -7.0 &-0.14  &-0.46  & -0.97\\
 12 & B8 & 4.93 & -7.4 &  250 & -7.0 &-0.05  &-0.34  & -0.57\\
 10 & A0 & 4.76 & -7.3 &  200 & -6.73 &+0.07  &-0.17  & -0.15\\
\hline
\end{tabular}
\end{center}
\end{table*}

\end{document}